\documentclass[pdflatex,sn-mathphys-num]{sn-jnl}

\usepackage{graphicx}%
\usepackage{multirow}%
\usepackage{amsmath,amssymb,amsfonts}%
\usepackage{amsthm}%
\usepackage{mathrsfs}%
\usepackage[title]{appendix}%
\usepackage{xcolor}%
\usepackage{textcomp}%
\usepackage{manyfoot}%
\usepackage{booktabs}%
\usepackage{algorithm}%
\usepackage{algorithmicx}%
\usepackage{algpseudocode}%
\usepackage{listings}%
\usepackage{color} 
\usepackage{soul}
\usepackage{units}
\usepackage{mathcomp}

\theoremstyle{thmstyleone}%
\theoremstyle{thmstyletwo}%

\theoremstyle{thmstylethree}%

\begin{document}

\title[Article Title]{Experimental assessment of a scalar-dissipation-rate sub-grid-scale closure for Large Eddy Simulation of turbulent premixed flames}


\author[]{\fnm{Morteza} \sur{Nahvi}}\email{morteza.nahvi@ubc.ca}
\author*[]{\fnm{Sina} \sur{Kheirkhah$^*$}}\email{sina.kheirkhah@ubc.ca}

\affil[]{\orgdiv{School of Engineering}, \orgname{The University of British Columbia}, \orgaddress{\street{1137 Alumni Ave.}, \city{Kelowna}, \postcode{V1V1V7}, \state{British Columbia}, \country{Canada}}}


\abstract{A sub-grid closure of the scalar-dissipation rate for the Large Eddy Simulation (LES) of turbulent premixed flames is assessed experimentally. Hotwire anemometry is performed to characterize the background turbulent flow, and the Rayleigh scattering measurements are conducted to acquire instantaneous temperature fields, which are used to obtain the scalar dissipation rate. Either methane or hydrogen-enriched (40\% by volume) methane is used as the fuel. The Karlovitz number varies between 1.7 and 34.5. The LES filter size is changed from 0.1 to 3.0 times the thermal thickness of the corresponding laminar flames. Both Gaussian and top-hat filtering methods are used. Statistical analyses are performed to investigate the characteristics of the measured, resolved, and unresolved scalar dissipation rate as well as how these are influenced by the normalized filter size, filter type, and the Karlovitz number. It is shown that the algebraic model allows for acceptable prediction of the Favre-filtered, temporally and spatially averaged scalar dissipation rate across the examined test conditions. The experimental measurements are used to gain insight into a key parameter used in the algebraic model. While differences between the previously proposed and our measured value of the model parameter for small Karlovitz number flames are reported, good agreement between our measured values and previously suggested values is obtained for relatively larger Karlovitz number flames. This finding is shown to be consistent across the examined filter sizes and filtering methods, despite the differences in the examined flame configuration of the present experimental study and those of past numerical investigations. This motivates future investigations to assess use of the algebraic model for simulating turbulent premixed flames with an extended range of operation.}

\keywords{Scalar dissipation rate modeling, Rayleigh scattering measurements, Large Eddy Simulations, Turbulent premixed flames}



\maketitle

\section{Introduction}\label{sec1}

The reaction rate ($\dot{\omega}$) is a key parameter in the Large Eddy Simulation (LES) of turbulent premixed flames~\mbox{\cite{selle2004compressible,jella2021attached,pitsch2006large}}. In the LES, the Sub-Grid Scale (SGS) models are usually developed using \textit{a-posteriori} analysis~\cite{langella2017large,langella2016large,lapenna2024posteriori}, \textit{a-priori} analysis~\cite{dunstan2013scalar,nilsson2019priori,kasten2022modeling} or both~\cite{proch2017flame,luo2021priori,wang2025priori,berger2025combustion} of the Direct Numerical Simulation (DNS) data. Compared to the analysis of the DNS data, the present study is motivated to assess a SGS model of $\dot{\omega}$ using experimental data.

Broadly, the flamelet-based and non-flamelet-based approaches are used to calculate $\dot{\omega}$ in the LES of turbulent premixed flames, see for example~\cite{pope1985flamelet,veynante2002turbulent,poinsot2005theoretical,pitsch2006large,cant2011rans}. The flamelet-based approaches generally used the thickened flame~\cite{colin2000thickened,han2019large}, the flame surface density~\cite{hawkes2011estimates}, flamelet generated manifolds~\cite{ho2023assessment}, and the level-set~\cite{moureau2009level} models. The non-flamelet-based approaches generally used the algebraic~\cite{dunstan2013scalar,gao2014algebraic}, the presumed Probability Density Function (PDF)~\cite{pfitzner2021new,liu2022machine,wang2022investigation}, the eddy-break-up~\cite{lysenko2018reynolds}, and moment closure~\cite{luo2021priori} models. Of interest to the present study is an algebraic model, which was originally developed by Bray \textit{et al.}~\cite{bray1979interaction,bray1980turbulent}, and the variations of this model were presented in several later investigations~\cite{kolla2009scalar,dunstan2013scalar,langella2017large}. For example, Kolla \textit{et al.}~\cite{kolla2009scalar} developed an algebraic model for the Favre-averaged Navier Stokes simulations. Later, Dunstan \textit{et al.}~\cite{dunstan2013scalar} and Langella \textit{et al.}~\cite{langella2017large} further developed the model of~\cite{kolla2009scalar} for the LES of turbulent premixed flames. In the algebraic models of~\cite{dunstan2013scalar,langella2017large}, the filtered reaction rate was calculated using
\begin{equation}
\label{Eq:omegadot}
    \overline{\dot{\omega}} = \frac{2\overline{\rho}\widetilde{N_c}}{2C_\mathrm{m}-1},
\end{equation}
where $\rho$ is the gas density, and $N_c =  \alpha (\nabla c)^2$ is the scalar dissipation rate. In Eq.~(\ref{Eq:omegadot}), $\alpha$ and $c$ are the thermal diffusivity and the temperature-based combustion progress variable, respectively. The over-bar symbol refers to the LES filtering; and, $\widetilde{\mathcal{X}} = \overline{\rho \mathcal{X}}/\overline{\rho}$, is the Favre-filtered $\mathcal{X}$, with $\mathcal{X}$ being a generic quantity of interest. Two types of filters, namely, Gaussian~\cite{langella2017large, dunstan2013scalar,gao2014algebraic,schumann2026priori} and box (or top-hat)~\cite{vreman2009subgrid,kazbekov2021physical} are of interest to the present study and could be used for the Favre-filtering purposes. The Gaussian filter attenuates the signal, leading to an underestimation of the Favre-filtered parameter~\cite{ranjan2016subgrid}; however, the box filter does not feature such a characteristic. In Eq.~(\ref{Eq:omegadot}), $C_\mathrm{m}$ is a modeling parameter which was originally introduced and defined by Bray~\cite{bray1979interaction,bray1980turbulent} and depends on $c$, the Probability Density Function (PDF) of $c$, and $\dot{\omega}$. Past studies~\cite{bray1979interaction,bray1980turbulent,chakraborty2011eeffects,ma2014validation,langella2017large} showed that depending on the above parameters, $C_\mathrm{m}$ varies between about 0.7 and 0.9.

The Favre-filtered scalar dissipation rate can be written as the summation of resolved ($\widetilde{\alpha}(\nabla \widetilde{c})^2$) and unresolved ($\widetilde{\epsilon_c}$) terms, given by~\cite{dunstan2013scalar,gao2014algebraic,ma2014validation,langella2017large}

\begin{equation}
\label{Eq:Ncterms}
\widetilde{N_c} = \widetilde{\alpha} (\nabla \widetilde{c})^2+\widetilde{\epsilon_c}.
\end{equation}
The label ``unresolved term'' contrasts the ``resolved term'', and is interchangeably used in the past literature to refer to the SGS scalar dissipation rate. Dunstan \textit{et al.}~\cite{dunstan2013scalar} studied the conditionally averaged (based on $\widetilde{c}$) values of $\widetilde{N_c}$. In their study~\cite{dunstan2013scalar}, the normalized LES filter size ($\Delta^+=\Delta/\delta_\mathrm{th}$, with $\Delta$ and $\delta_\mathrm{th}$ being the LES filter size and the thermal flame thickness, respectively) varied between 0.2 and 3.0. The thermal flame thickness was defined~\cite{dunstan2013scalar} as the ratio of the difference between the adiabatic flame temperature and the unburned gas temperature to the maximum absolute value of the temperature gradient for the corresponding laminar flame. For all examined $\Delta^+$ values, their results~\cite{dunstan2013scalar} showed that the variation of the conditionally averaged $\widetilde{N_c}$ versus $\widetilde{c}$ features a parabolic-like distribution. It was presented in~\cite{dunstan2013scalar} that increasing $\Delta^+$ from 0.2 to 3.0 decreased the maximum value of the conditionally averaged $\widetilde{N_c}\delta_\mathrm{th}/S_\mathrm{L,0}$ (with $S_\mathrm{L,0}$ being the unstrecthed laminar flame speed) from about 1.6 to 0.4. Also, increasing $\Delta^+$ from 0.2 to 3.0 decreased the value of $\widetilde{c}$ at which $\widetilde{N_c}\delta_\mathrm{th}/S_\mathrm{L,0}$ maximized from about 0.7 to 0.5~\cite{dunstan2013scalar}. It was discussed in~\cite{dunstan2013scalar} that increasing the filter size increased the number of data points associated with fresh reactants and fully burnt gas (both of which featuring close to zero values of $\widetilde{N_c}\delta_\mathrm{th}/S_\mathrm{L,0}$) within the filter, and this led to the reported smaller values of $\widetilde{N_c}\delta_\mathrm{th}/S_\mathrm{L,0}$ at larger normalized filter sizes. 

Kasten \textit{et al.}~\mbox{\cite{kasten2022modeling}} investigated the conditionally averaged values of $\widetilde{\epsilon_c}\delta_\mathrm{th}/S_\mathrm{L,0}$ filtering the DNS results of turbulent premixed flames. Their results~\mbox{\cite{kasten2022modeling}} showed that the variation of $\widetilde{\epsilon_c}\delta_\mathrm{th}/ S_\mathrm{L,0}$ versus $\widetilde{c}$ features both positive and negative values for filter sizes smaller than $\delta_\mathrm{th}$. Their results~\cite{kasten2022modeling} showed that, for $\Delta^+ =0.8$, the smallest values of $\widetilde{\epsilon_c}\delta_\mathrm{th}/S_\mathrm{L,0}$ occurred at $\widetilde{c} \approx 0.7$. However, compared to the results for small filter sizes, the conditionally averaged values of $\widetilde{\epsilon_c}\delta_\mathrm{th}/S_\mathrm{L,0}$ remained positive for filter sizes larger than $\delta_\mathrm{th}$. Their results~\cite{kasten2022modeling} showed that increasing $\Delta^+$ from 0.8 to 5.6 increased the value of $\widetilde{c}$ at which $\widetilde{\epsilon_c}\delta_\mathrm{th}/ S_\mathrm{L,0}$ featured a maximum from about 0.3 to 0.5, with this trend being independent of the examined test condition in~\cite{kasten2022modeling}.

An algebraic model of $\widetilde{\epsilon_c}$ for LES of the turbulent premixed flames was given by~\cite{dunstan2013scalar}
\begin{equation}
\label{Eq:epsilon}
\widetilde{\epsilon_c} = \left\{ \frac{\mathcal{F}}{\beta_c}\left[2K_c\frac{S_\mathrm{L,0}}{\delta_\mathrm{th}}+(C_3-\tau C_4 Da_\Delta)(\frac{2u'_\Delta}{3\Delta})\right]\right\}\widetilde{c}(1-\widetilde{c}).
\end{equation}
Equation~(\ref{Eq:epsilon}) suggests that $\widetilde{\epsilon_c}$ is linearly related to $\widetilde{c}(1-\widetilde{c})$, with the slope of the linear relation being presented by the term in the curly braces. In Eq.~(\ref{Eq:epsilon}), $K_c = 0.79\tau$ \cite{kolla2009scalar,dunstan2013scalar}, with $\tau$ being the ratio of the burnt and unburnt gas temperatures minus unity. In Eq.~(\ref{Eq:epsilon}), $\mathcal{F} = 1-\exp(-3\Delta^+/4)$, $Da_\Delta = (u'_\Delta/S_\mathrm{L,0})^{-1}(\Delta/\delta_\mathrm{th})$, $C_3 = 1.5\sqrt{Ka_\Delta}/(1+\sqrt{Ka_\Delta})$, and $C_4 = 1.1/(1+Ka_\Delta)^{0.4}$, with $Ka_\Delta = (u'_\Delta/S_\mathrm{L,0})^{3/2}(\Delta/\delta_\mathrm{th})^{-1/2}$. In the Favre-averaged Navier Stokes framework, Kolla \textit{et al.}~\cite{kolla2009scalar} suggested a value of 6.7 for $\beta_c$. However, for LES, Dunstan~\textit{et al.}~\cite{dunstan2013scalar} and Langella \textit{et al.}~\cite{langella2017large} used values of 2.4 and 7.5, respectively. Aiming to improve the predictions of LES, Ma \textit{et al.}~\cite{ma2014validation} analyzed their DNS data and suggested a formulation that related $\beta_c$ to $\tau$ and $C_\mathrm{m}$. They~\cite{ma2014validation} concluded that, although their proposed formulation led to results that compared well with those of the experiments reported in~\cite{nguyen2009experimental,sjunnesson1992cars}, alternative values of $\beta_c$ would yield further improvements to the LES of the turbulent premixed flames. Addressing this, a dynamic model of $\beta_c$ was formulated later and presented in the LES of turbulent premixed flames in~\cite{ferrante2024dynamic}. There~\cite{ferrante2024dynamic}, significant probability of $0.1 \lesssim \beta_c \lesssim 100$ was reported. Although past numerical investigations have extensively studied $\beta_c$, this parameter has not been directly obtained using experimental data to our best knowledge.

In Eq.~(\ref{Eq:epsilon}), $u'_\Delta$ (the root-mean-square of the SGS velocity fluctuations~\cite{chakraborty2007influence,dunstan2013scalar}) is an important parameter for estimating $\widetilde{\epsilon_c}$. Langella \textit{et al.}~\cite{langella2017large} performed DNS of turbulent premixed Bunsen flames and showed $u'_\Delta$ remains nearly constant for $0.1 \lesssim \widetilde{c} \lesssim 0.9$ for a given LES filter size and fixed background flow condition. For example, changing $\widetilde{c}$ from 0.1 to 0.9 varied $u'_\Delta$ (obtained from filtering their~\cite{langella2017large} DNS data) from about $1.3S_\mathrm{L,0}$ to $0.9S_\mathrm{L,0}$ and for $\Delta^+ = 0.8$. For the same background flow condition but for a larger normalized filter size of $\Delta^+ = 2.8$, increasing $\widetilde{c}$ from 0.1 to 0.9 changed $u'_\Delta$ from about $3.1S_\mathrm{L,0}$ to $2.5S_\mathrm{L,0}$~\cite{langella2017large}. In that study~\cite{langella2017large}, in addition to filtering the DNS data, $u'_{\Delta}$ was modeled using the velocity scale-similarity~\cite{pope2001turbulent}, kinetic-energy scale-similarity~\cite{ferziger1997large}, a shear-stress closure~\cite{lilly1966representation}, as well as a modified by~\cite{langella2017large} version of the Smagorinsky's model in~\cite{pope2001turbulent}. Comparing with their filtered DNS, Langella \textit{et al.}~\cite{langella2017large} discussed that $u'_\Delta$ obtained from~\cite{pope2001turbulent,lilly1966representation,ferziger1997large} underestimated the conditionally averaged  $\widetilde{\epsilon_c}\delta_\mathrm{th}/S_\mathrm{L,0}$; however, the modified by~\cite{langella2017large} version of the Smagorisnky model led to prediction of $\widetilde{\epsilon_c}\delta_\mathrm{th}/S_\mathrm{L,0}$ closer to those of their filtered DNS. 

Imperative to assessing the parameters in Eq.~(\ref{Eq:epsilon}) is the type of the LES filtering used for calculating $\widetilde{\epsilon_c}$ and $\widetilde{c}$. It is noted that past studies which investigated $\widetilde{N_c}$, $\widetilde{\epsilon_c}$, and the variables in the curly braces of Eq.~(\ref{Eq:epsilon}) used the Gaussian type filter in their analyses~\cite{dunstan2013scalar,gao2014algebraic,langella2017large,kasten2022modeling}. While it has been shown~\cite{kazbekov2021physical} that changing the filter type from Gaussian to top-hat (referred to as ``box'' here) may have an impact on the terms in the transport equation of the turbulent kinetic energy of premixed flames, the effects of the filter type on the scalar dissipation rate and the corresponding algebraic model remain to be investigated. Specifically, the effects of changing the filter type from Gaussian (used in~\cite{dunstan2013scalar,gao2014algebraic,langella2017large,kasten2022modeling}) to box (see that in~\cite{vreman2009subgrid,hernandez2011laboratory,shahbazian2015subfilter}) on parameters relevant to the algebraic model of the scalar dissipation rate, e.g., $\beta_c$, in Eq.~(\ref{Eq:epsilon}), remain to be investigated.

Along with the numerical investigations to study the scalar dissipation rate and its SGS model, experimental investigations of the scalar dissipation rate for non-reacting flows (e.g., \cite{soulopoulos2014scalar,stetsyuk2016scalar}), turbulent non-premixed flames (e.g., \cite{masri2004compositional,kaiser2007imaging,sutton2013measurements}), and turbulent premixed flames~(e.g., \cite{o1997scalar,trueba2023flame,skiba2022premixed,kamal2017scalar,mohammadnejad2024effects,mohammadnejad2025spectral}) have been performed in the past, with the latter group of investigations being of relevance to the present study. It was experimentally shown~\cite{o1997scalar,trueba2023flame,skiba2022premixed} that the conditionally averaged scalar dissipation rate follows a parabolic-like distribution with the progress variable. Also, past experimental studies~\cite{trueba2023flame,skiba2022premixed,kamal2017scalar,mohammadnejad2024effects} obtained $C_\mathrm{m}$ for several background flow conditions. The findings of~\cite{trueba2023flame,skiba2022premixed,kamal2017scalar} suggested that $C_\mathrm{m}$ depends on the gas temperature; and later, Mohammadnejad and Kheirkhah~\cite{mohammadnejad2024effects} quantified that moving form the fresh gas to the hot combustion products increased $C_\mathrm{m}$ by a few folds. Recently, Mohammadnejad and Kheirkhah~\cite{mohammadnejad2025spectral} experimentally estimated the burning velocity of turbulent premixed flames using the scalar dissipation rate and investigated the influence of the background turbulence on the spectral characteristics of both the scalar dissipation rate and the related burning velocity.

Although past experimental investigations that assessed the scalar dissipation rate are of significant importance and of relevance to the present study, their analyses were performed in the Reynolds- or Favre-averaged Navier Stokes frameworks. To our best knowledge, no investigation has experimentally examined the filtered scalar dissipation rate in the framework of LES and measured the SGS scalar dissipation rate yet. The first objective of the present study is to experimentally measure $\widetilde{N_c}$ and $\widetilde{\epsilon_c}$ as well as to assess how these parameters are influenced by the LES filter size and type for several background turbulent flow conditions. The second objective of the present study is to experimentally assess the algebraic model in Eq.~(\ref{Eq:epsilon}) and estimate $\beta_c$ in that equation.

\section{Experimental methodology}\label{sec 2}

The experimental methodology is identical to that used in~\cite{mohammadnejad2024effects}. Here, a summary is provided and the reader is referred to \cite{mohammadnejad2024effects} for further details. The experimental setup was composed of a diffuser section, a settling chamber, a contraction section, a nozzle, and a flame-holder to produce turbulent premixed V-shaped flames. The exit diameter of the nozzle section was \unit[47.7]{mm} and the diameter of the flame-holder was \unit[4.0]{mm}. The distance between the axis of the flame-holder and the exit plane of the nozzle was~\unit[5.0]{mm}. The air flow was provided using a rotary screw compressor (model GA37FF from Atlas Copco), and the air flowrate was measured using an ALICAT MCRH~5000. The fuel was either methane or 60\% by volume methane mixed with 40\% by volume hydrogen. The methane and hydrogen-enriched methane were provided from separate bottles with chemical purities of 99\% and 99.999\%, respectively. The methane and hydrogen-enriched methane flowrates were measured using SLA5853 mass flow controller from Brooks. Either two perforated plates or an active turbulence generator were used, with details of both turbulence generating mechanisms discussed in~\cite{mohammadnejad2022new} and are not presented here for brevity.

The hotwire anemometry (HWA) was used for characterizing the background cold flow. For such characterization, the flame-holder was removed. The location of the HWA was \unit[37.0]{mm} downstream of the burner exit plane and at the burner centerline. The wire used for HWA was $\unit[5]{\tcmu m}$ in diameter and had a measurement length of $l_\mathrm{w} =$\unit[1.25]{mm}. The probe temperature was kept constant at an overheat ratio of 0.7. The acquisition frequency of HWA was \unit[10]{kHz}, and the data was acquired for \unit[90]{s}. A Nd:YAG laser and a harmonic generator were used to produce a \unit[532]{nm} beam. The energy of the beam was about \unit[470.0]{mJ} per pulse with a shot-to-shot variation of about \unit[4]{mJ}. The laser beam was converted to a $\unit[216]{\tcmu m}$ laser sheet using optics, with details provided in \cite{mohammadnejad2024effects}. The Rayleigh scattering data was acquired using an sCMOS camera, which was equipped with a bandpass filter (center wavelength of \unit[532]{nm} and full-width at half maximum of \unit[20]{nm}). The camera's physical sensor size and the camera's pixel resolution were 20.0 and $\unit[39.6]{\tcmu m}$, respectively. The field-of-view of the captured Rayleigh scattering signal was defined in a Cartesian coordinate system. The $y-$axis was along the nozzle centerline, and the $x-y$ plane coincided with the plane of Rayleigh scattering measurements. The origin of the coordinate system was at the exit plane of the nozzle. For each examined test condition (discussed later), 520 Rayleigh scattering images of the reacting flow were acquired.

\begin{figure}[h]
\centerline{\includegraphics[width=1\textwidth]{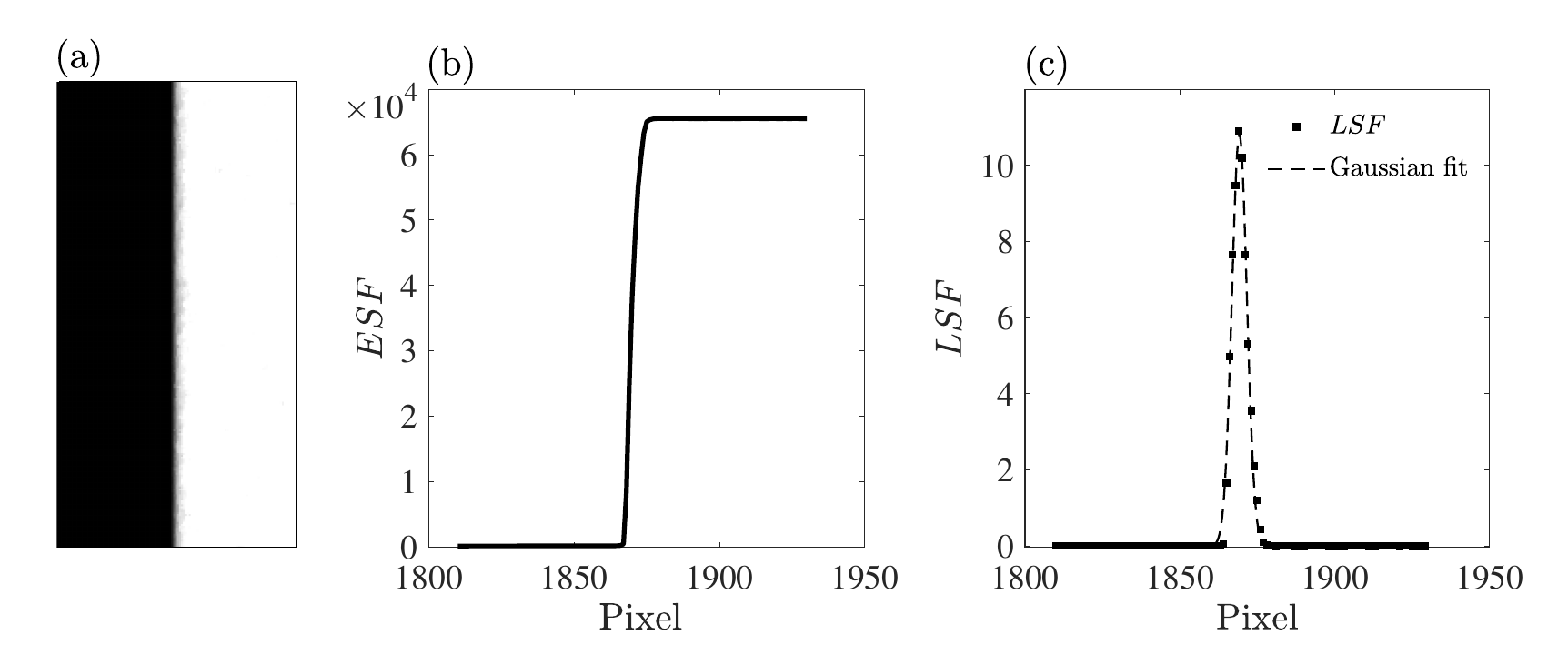}}
\caption{(a) is the image of a sharp edge. (b) is the variation of average intensity along a line perpendicular to the edge. (c) is the LSF (see the solid square data points) and a Gaussian variation fit to the LSF (see the dashed curve).}\label{Fig:LSF}
\end{figure}

The resolving limit of the Rayleigh scattering measurements was the larger of the laser sheet thickness and the resolving limit of imaging obtained from a USAF 1951 target plate. To this end, a small window of a USAF 1951 target plate, with a sample shown in Fig.~\ref{Fig:LSF}(a), was used. Then, the Edge Spread Function (ESF), which is the vertically-averaged intensity, versus the horizontal axis (in pixels) perpendicular to the edge was obtained and presented in Fig.~\ref{Fig:LSF}(b). The derivative of ESF with respect to the horizontal axis is referred to as the Line Spread Function (LSF), which is presented in Fig.~\ref{Fig:LSF}(c). A Gaussian variation was fit to the LSF, with this fit shown by the dotted curve in the figure. The Full Width at Half Maximum (FWHM) of the Gaussian fit was calculated and was approximately $\unit[198]{\tcmu m}$, which is smaller than the laser sheet thickness; and as a result, the resolving limit of our imaging was $\unit[216]{\tcmu m}$. In our earlier study~\cite{mohammadnejad2024effects}, the normalized energy and dissipation spectra of the temperature fluctuations was used and it was discussed that a spatial resolution of at least $\unit[314]{\tcmu m}$ is required to resolve the temperature fluctuations. This resolving limit is larger than the resolution of our Rayleigh scattering measurements.
 
A summary of the examined test conditions is presented in Table~\ref{tab:Conditions}. In the table, the generic forms of U$X$F$Y$H$Z$P2 and U$X$F$Y$H$Z$AT are used to identify the test conditions for which two perforated plates or the active turbulence generator were used, respectively. In the symbols used for identifying the test conditions, $X$ and 0.$Y$ are the examined mean bulk flow velocity ($U$) and the fuel-air equivalence ratio ($\phi$), respectively. $Z$ is the volumetric percentage of the hydrogen-enrichment ($\mathrm{H_2}\%$). In the table, $L$ is the integral length scale and is estimated using the auto-correlation of the velocity data, with the details of the calculation provided in~\cite{mohammadnejad2022new}. The Taylor and Kolmogorov length scales were obtained using $l_\mathrm{T}=LRe^{-1/2}_\mathrm{T}$ and $\eta=LRe^{-3/4}_\mathrm{T}$, respectively. The turbulent Reynolds number is estimated from $Re_\mathrm{T} = u'L/\nu$, with $u'$ and $\nu$ being the RMS velocity fluctuations of the background turbulent cold flow and the reactants kinematic viscosity at \unit[300]{K}, respectively. In Table~\ref{tab:Conditions}, $Da = (u'/S_\mathrm{L,0})^{-1}(L/\delta_\mathrm{L})$ and $Ka = (u'/S_\mathrm{L,0})^{3/2}(L/\delta_\mathrm{L})^{-1/2}$ are the Damk\"{o}hlder and Karlovitz numbers, respectively, with $\delta_\mathrm{L}$ being the laminar flame thickness. The unstretched laminar flame speed was calculated from the Cantera~\cite{cantera} simulations of one-dimensional freely propagating flames for the examined fuel and air mixtures and using GRI-Mech 3.0~\cite{smith1999gri}. The laminar flame thickness was estimated from $\delta_\mathrm{L} = (\lambda/c_\mathrm{P})/(\rho_\mathrm{u}S_\mathrm{L,0})$. In this calculation and similar to past studies \cite{mohammadnejad2020thick,driscoll2020premixed,mohammadnejad2021contributions}, the reactants density ($\rho_\mathrm{u}$) was estimated at \unit[300]{K}, while the reactants thermal conductivity ($\lambda$) and heat-capacity-at-constant-pressure ($c_\mathrm{P}$) were both evaluated at \unit[1500]{K}, using \cite{mathur1967thermal}. For the combination of the examined $\phi$ and $\mathrm{H_2}\%$, the effective Lewis number of the examined methane-air and hydrogen-enriched methane air flames were calculated in \cite{mohammadnejad2024effects} and are about 1.0 and 0.8, respectively. It is important to note that, while the combination of $\phi$ and $\mathrm{H_2\%}$ in the present study is such that the Lewis number is rather close to unity, both $S_\mathrm{L,0}$ and $\delta_\mathrm{L}$ change notably. Specifically, $S_\mathrm{L,0}$ and $\delta_\mathrm{L}$ are \unit[0.19]{m/s} and \unit[0.33]{mm} for the examined methane-air flames and \unit[0.08]{m/s} and \unit[0.89]{mm} for the hydrogen-enriched methane air flames. Given these values and the characteristics of the examined background turbulent flow, the turbulent methane-air premixed flames featured $1.7\leq Ka \leq 5.8$; and, the turbulent hydrogen-enriched methane-air flames featured $11.2\leq Ka \leq 34.5$, as listed in Table~\ref{tab:Conditions}. It is important to note that $\delta_\mathrm{L}$ is used for calculating several non-dimensional parameters, e.g., $Ka$ and $Da$ as listed in Table~\ref{tab:Conditions} of the present study and similar to past investigations reviewed in~\cite{driscoll2020premixed}. Compared to $\delta_\mathrm{L}$, however, $\delta_\mathrm{th}$ is used for normalizing the LES filter size and estimating several parameters relevant to Eq.~(\ref{Eq:epsilon}), e.g., $Ka_\Delta$ and $Da_\Delta$. The Cantera~\cite{cantera} simulations (with GRI-Mech 3.0~\cite{smith1999gri}) of freely propagating flames were used to estimate $\delta_\mathrm{th}$, which was \unit[0.66]{mm} and \unit[1.29]{mm} for the examined methane-air and hydrogen-enriched methane-air flames, respectively.

\begin{table*}[!htbp]
	\caption{Tested experimental conditions. $U$ and $u'$ are in m/s; and, $L$, $l_\mathrm{T}$, and $\eta$ are in~mm. $S_\mathrm{L,0}=\unit[0.19]{m/s}$ ($S_\mathrm{L,0}=\unit[0.08]{m/s}$) and $\delta_\mathrm{L}=\unit[0.33]{mm}$ ($\delta_\mathrm{L}=\unit[0.89]{mm}$) for the conditions with $\phi=0.7$ ($\phi=0.5$) and $\mathrm{H_2}\% = 0\%$ ($\mathrm{H_2}\% = 40\%$).}
	\label{tab:Conditions}
	\centering
    \scalebox{0.96}{
    \makebox[\textwidth][c]{%
		\begin{tabular}{l|ccccccccccc}
			\hline
			\hline
			Conditions & $U$ & $\phi$ & $\mathrm{H_2}\%$ & $u'$ & $L$ & $l_\mathrm{T}$ & $\eta$ & $L/\delta_\mathrm{L}$ & $Re_\mathrm{T}$ & $Da$ & $Ka$\\
			\hline
			U5F7H00P2 & 5.0 & 0.7 & 0 & 0.79 & 7.2 & 0.38 & 0.09 & 21.8 & 350 & 5.4 & 1.7\\
			U5F5H40P2 & 5.0 & 0.5 & 40 & 0.79 & 7.2 & 0.39 & 0.09 & 8.0 & 343 & 0.8 & 11.2\\
			U7F7H00P2 &7.0 & 0.7 & 0 & 1.16 & 8.4 & 0.34 & 0.07 & 25.5 & 603 & 4.3 & 2.9\\
			U7F5H40P2 & 7.0 & 0.5 & 40 & 1.16 & 8.4 & 0.34 & 0.07 & 9.4 & 591 & 0.6 & 18.7\\
			U5F7H00AT & 5.0 & 0.7 & 0 & 1.23 & 10.8 & 0.38 & 0.07 & 32.8 & 820 & 5.2 & 2.8\\
			U5F5H40AT & 5.0 & 0.5 & 40 & 1.23 & 10.8 & 0.38 & 0.07 & 12.1 & 804 & 0.8 & 17.8\\
			U7F7H00AT & 7.0 & 0.7 & 0 & 1.84 & 9.8 & 0.29 & 0.05 & 29.7 & 1115 & 3.1 & 5.3\\
			U7F5H40AT & 7.0 & 0.5 & 40 & 1.84 & 9.8 & 0.30 & 0.05 & 11.0 & 1093 & 0.5 & 34.5\\
			\hline
			\hline
	\end{tabular}}}
	\vspace{-3mm}
\end{table*}

\section{Data reduction}\label{sec 3}

The Rayleigh scattering images were used to obtain the gas temperature from~\mbox{\cite{skiba2022premixed}}
\begin{equation}
\label{Eq:RS}
T = T_{\mathrm{ref}}
\frac{I_\mathrm{i}}{I_{\mathrm{i,ref}}}
\frac{I_{\mathrm{R,ref}}}{I_\mathrm{R}}
\frac{\sigma_{\mathrm{mix}}}{\sigma_{\mathrm{mix,ref}}}.
\end{equation}
In \mbox{Eq.~(\ref{Eq:RS})}, ``ref'' stands for the reference condition, corresponding to air at \mbox{\( T_{\mathrm{ref}} = 300~\mathrm{K} \)}. Considering the small standard deviation of the laser energy fluctuations (which was less than 1\% of the mean value), $I_\mathrm{i} / I_{\mathrm{i,ref}} \approx 1$. In Eq.~(\ref{Eq:RS}), $I_\mathrm{R}$ and $ I_{\mathrm{R,ref}}$ represent the Rayleigh scattering signals for the reacting and reference conditions, respectively. Both $I_\mathrm{R}$ and $I_{\mathrm{R,ref}}$ were corrected for background and noise. In Eq.~(\ref{Eq:RS}), $\sigma_{\mathrm{mix}}$ and $\sigma_{\mathrm{mix,ref}}$ are the Rayleigh scattering cross-sections of the reacting mixture and air at a laser excitation wavelength of \unit[532]{nm}, respectively.

Equation~(\ref{Eq:RS}) along with the Cantera~\cite{cantera} simulations of the corresponding laminar premixed freely propagating flames (with GRI-Mech 3.0~\cite{smith1999gri}) were used to estimate the normalized Rayleigh scattering intensity $I_\mathrm{R}/I_{\mathrm{R,ref}}$. The correlations between gas temperature and the normalized Rayleigh scattering intensity were derived from these Cantera simulations and discussed for both fuel compositions in~\cite{mohammadnejad2024effects}. These correlations were used to determine coefficients of fitting relations, following the methodology in~\cite{skiba2022premixed}. The resulting fitted curves were then employed to obtain the temperature fields of the turbulent premixed flames. The relationship between gas temperature and $I_\mathrm{R}/I_{\mathrm{R,ref}}$ (obtained from Cantera simulations) was also examined for different stretch factors and chemical kinetic mechanisms. The results suggested that changing the stretch factor from 0.1 to 10 did not change the above relationship using either GRI-Mech 3.0~\cite{smith1999gri}~(as also reported in \cite{mohammadnejad2024effects}) or the FFCM-2 mechanism~\cite{ZDV2023}. The above analysis suggests that the relation between gas temperature and $I_\mathrm{R}/I_{\mathrm{R,ref}}$ is nearly insensitive to the stretch effects and the utilized chemical kinetic mechanisms for the mixtures examined in the present study.

\begin{figure}[h!]
	\centerline{\includegraphics[width=1.0\textwidth]{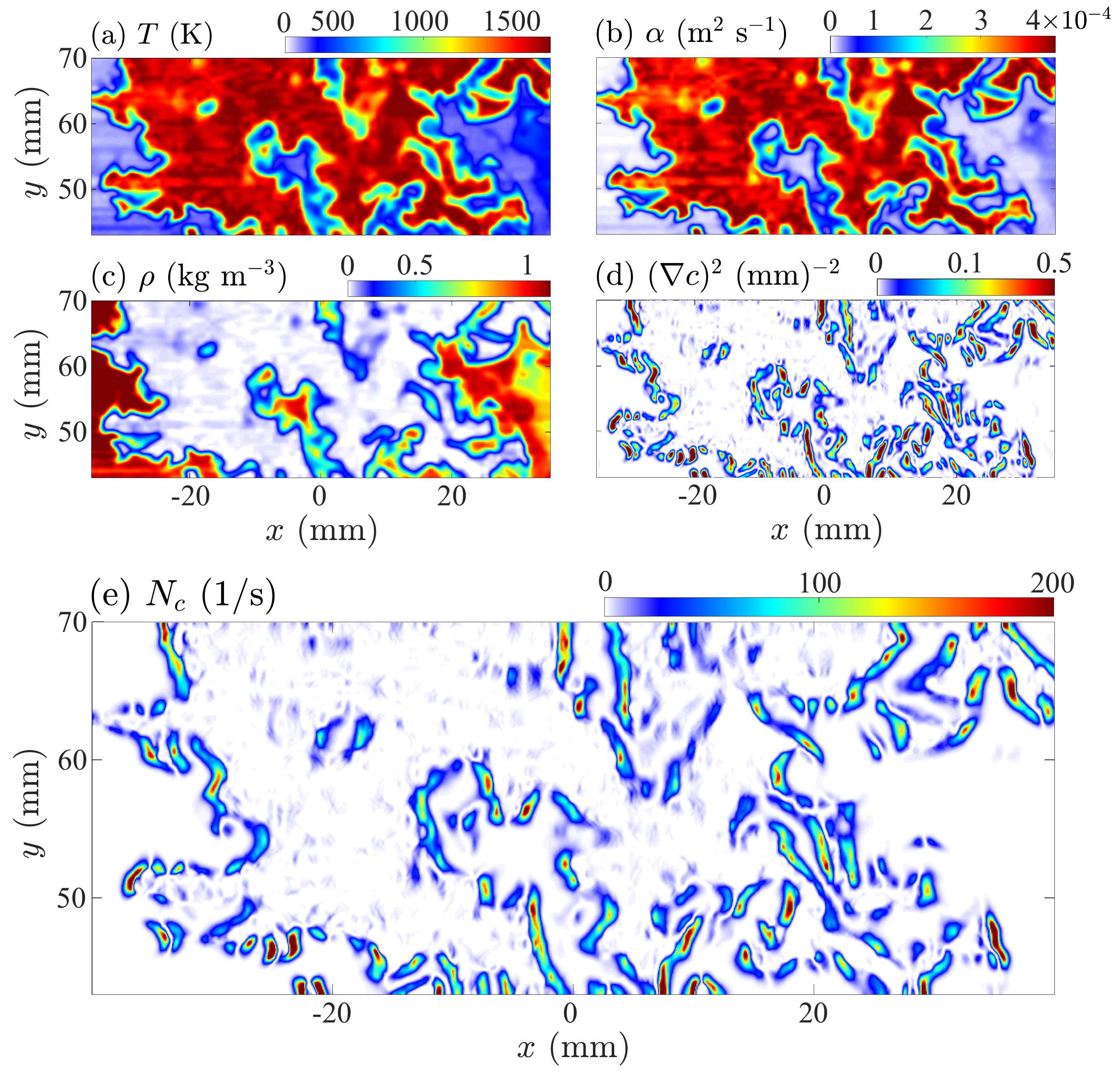}}
	\caption{Instantaneous variations of (a) $T$, (b) $\alpha$, (c) $\rho$, (d) $(\nabla c)^2$, and (e) $N_c$. The presented fields correspond to the test condition with $Ka = 34.5$.}
	\label{Fig:datareduction}
\end{figure}

The two-dimensional temperature fields were filtered using wavelet-based denoising, self-guided edge-preserving, and 5$\times$5 median-based filters to remove ``salt and paper'' noise, with an example of a temperature field presented in Fig.~\ref{Fig:datareduction}(a). The Rayleigh scattering Region-of-Interest (ROI) extended from $x =  \unit[-35.0]{mm}$ to \unit[35.0]{mm} and from $y =  \unit[43.0]{mm}$ to \unit[70.0]{mm}. The signal-to-noise ratio (SNR) of the temperature field was obtained via calculating the ratio of the RMS temperature fluctuations divided by the mean temperature inside a window of \unit[1]{mm}$\times$\unit[1]{mm} in the products, similar to~\cite{trueba2023flame}. Across the examined test conditions, the SNR varied
between 100 and 150, with these values being comparable to those reported in past studies, see for example~\cite{trueba2023flame}.

Assessment of the resolved and unresolved terms in Eq.~(\ref{Eq:Ncterms}) requires estimating the density, thermal diffusivity, and scalar dissipation rate. The Cantera~\cite{cantera} simulations (with GRI-Mech 3.0~\cite{smith1999gri}) of the unstretched laminar premixed flames were used to obtain the relations between temperature and thermal diffusivity as well as temperature and density. Then, the temperature fields along with these relations were used to obtain the two-dimensional variations of $\alpha$ and $\rho$. For example, the variations of $\alpha$ and $\rho$ for the temperature field shown in Fig.~\ref{Fig:datareduction}(a) are presented in Figs.~\ref{Fig:datareduction}(b)~and~(c), respectively. The temperature fields were used to calculate the combustion progress variable and $(\nabla c)^2$. It is acknowledged that calculating the latter requires estimating $\partial c/\partial x$ and $\partial c/\partial y$ as well as the derivative of the progress variable with respect to the direction normal to the plane of Rayleigh scattering measurements. This last derivative was not available in the present study, and as such $(\nabla c)^2$ is approximated as $(\nabla c)^2 \approx (\partial c/\partial x)^2 + (\partial c/\partial y)^2$, with representative results corresponding to the temperature filed in Fig.~\ref{Fig:datareduction}(a) presented in Fig.~\ref{Fig:datareduction}(d). Such two-dimensional (2D) approximation of $(\nabla c)^2$ has implications for the assessment of the terms in the algebraic model in Eq.~(\ref{Eq:epsilon}), with details discussed later in section~\ref{sec 4}. Finally, the thermal diffusivity and $(\nabla c)^2$, see those in Figs.~\ref{Fig:datareduction}(b) and (d), were used to calculate $N_c$. The scalar dissipation rate field corresponding to the results in Figs.~\ref{Fig:datareduction}(a--d) is shown in Fig.~\ref{Fig:datareduction}(e).

Several 2D fields, e.g., the scalar dissipation rate, density, and thermal diffusivity, were separately filtered using the Gaussian and the sharp cut-off box filters. For the Gaussian filtering method, $\Delta$ was the kernel size (i.e., the standard deviation of the Gaussian distribution) following the approach in past studies~\cite{langella2017large,dunstan2013scalar,gao2014algebraic,schumann2026priori}. For the box filtering method, the filtering kernel, i.e. $\kappa(\Delta)$, in 2D space was~\cite{chen2005implication}

\begin{equation}
\kappa_{\Delta}(\mathbf{x} - \mathbf{r}) =
\begin{cases}
\dfrac{1}{\Delta^2}, & \text{if } |\mathbf{x} - \mathbf{r}| \leq \dfrac{\Delta}{2}, \\[8pt]
0, & \text{otherwise,}
\end{cases}
\end{equation}
where $\mathbf{x} - \mathbf{r}$ was the vector connecting the center of the filtering box (here filtering window) to an arbitrary point in the 2D space.

\section{Results}\label{sec 4}
The results are grouped into two subsections. In the first subsection, the characteristics of the scalar dissipation rate along with its resolved and unresolved terms (see, Eq.~(\ref{Eq:Ncterms})) are discussed. In the second subsection, the algebraic model and constituent terms (see, Eq.~(\ref{Eq:epsilon})) are assessed.

\subsection{Characteristics of $\widetilde{N_c}$, $\widetilde{\alpha}(\nabla \widetilde{c})^2$, and $\widetilde{\epsilon_c}$}\label{subsec:sdr_characteristics}
\begin{figure}[h!]
\centerline{\includegraphics[width=1\textwidth]{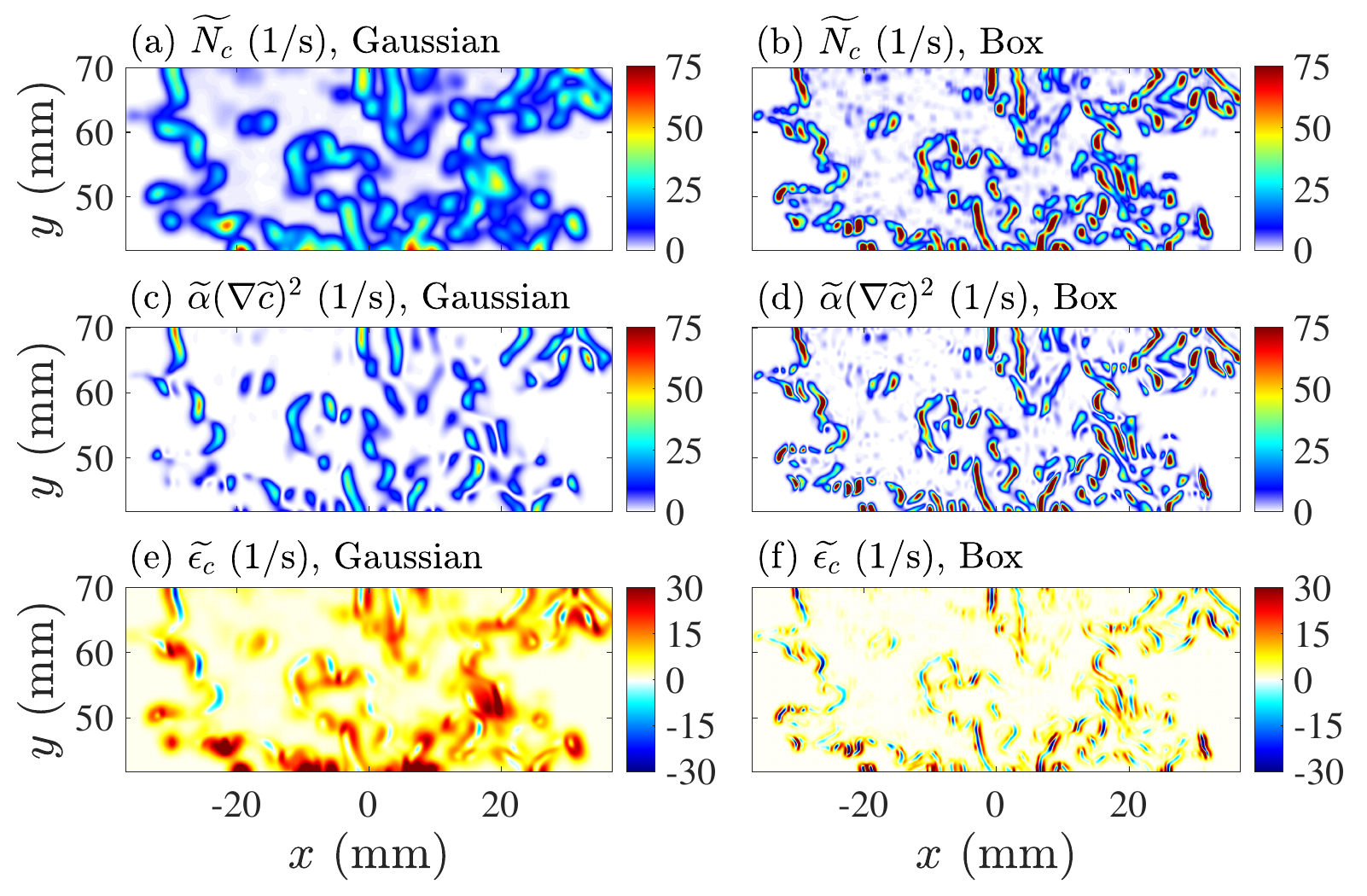}}
\caption{Instantaneous (a) and (b) $\widetilde{N_c}$, (c) and (d) $\widetilde{\alpha}(\nabla \widetilde{c})^2$, and (e) and (f) $\widetilde{\epsilon_c}$ obtained from the Gaussian filtering (left column) and box filtering (right column) methods. The results correspond to test condition U7F5H40AT ($Ka = 34.5$) and for $\Delta^+ = 1.0$.}\label{Fig:Resolvedandunresolved}
\end{figure}

\begin{figure}[h!]
\centerline{\includegraphics[width=1.0\textwidth]{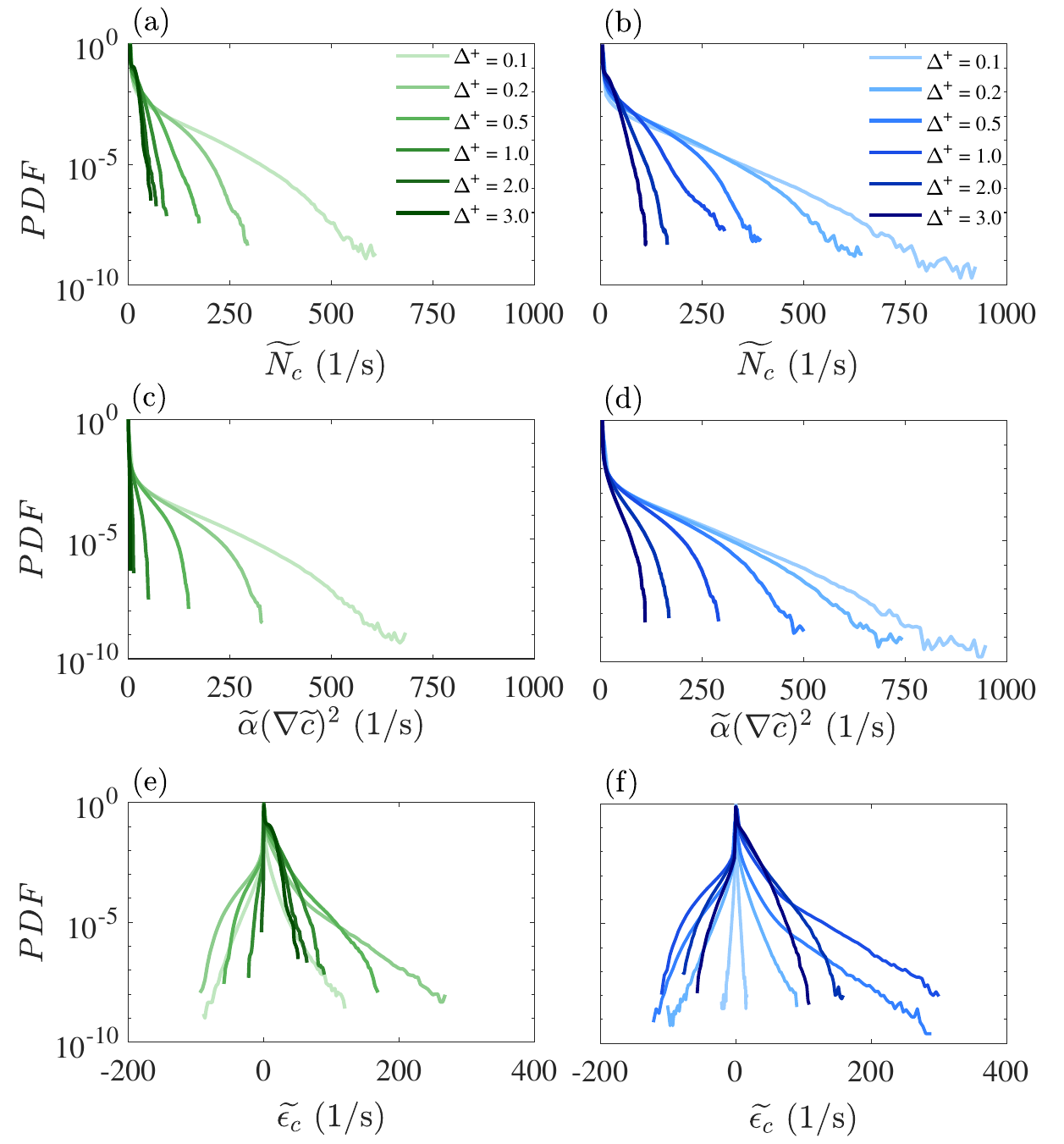}}
\caption{PDFs of (a) and (b) $\widetilde{N_c}$, (c) and (d) $\widetilde{\alpha}(\nabla \widetilde{c})^2$, and (e) and (f) $\widetilde{\epsilon_c}$ for Gaussian filtering (left column, in shades of green) and box filtering (right column, in shades of blue) for test condition U7F5H40AT ($Ka = 34.5$). The filter size ranges from $\Delta^+ = 0.1$ (light color) to $\Delta^+ = 3.0$ (dark color).}\label{Fig:NCPDF}
\end{figure}

The filtered density ($\overline{\rho}$) and filtered multiplication of density and scalar dissipation rate ($\overline{\rho N_c}$) fields were used to examine the unresolved and resolved terms in Eq.~(\ref{Eq:Ncterms}). Examples of $\widetilde{N_c}$, $\widetilde{\alpha}(\nabla \widetilde{c})^2$, and $\widetilde{\epsilon_c}$ are presented in Fig.~\ref{Fig:Resolvedandunresolved} for the Gaussian (left column) and box (right column) filtering methods. The presented fields correspond to the test condition and data shown in Fig.~\ref{Fig:datareduction} and for the normalized filter size of $\Delta^+ = 1$. To facilitate comparisons between the results of the Gaussian and box filtering methods, the maximum values of the results on the left panels were kept to be the same as those on the right panels in Fig.~\ref{Fig:Resolvedandunresolved}. This led to a slight saturation in the presentation of the results corresponding to the box filtering method. Comparing the results in Fig.~\ref{Fig:Resolvedandunresolved}(a, c, and e) with those in Fig.~\ref{Fig:Resolvedandunresolved}(b, d, and f), it can be seen that $\widetilde{N_c}$, $\widetilde{\alpha}(\nabla \widetilde{c})^2$, and $\widetilde{\epsilon_c}$ obtained from the Gaussian filtering method feature relatively broader structures compared to those obtained from the box filtering method. It is also observed that $\widetilde{\epsilon_c}$ obtained from both filtering methods features both positive and negative values. These observations were further analyzed using the normalized-by-maximum PDFs of $\widetilde{N_c}$, $\widetilde{\alpha}(\nabla \widetilde{c})^2$, and $\widetilde{\epsilon_c}$ for all examined filter sizes, with the results for the test condition with the largest Karlovitz number (U7F5H40AT, $Ka = 34.5$) shown in Figs.~\ref{Fig:NCPDF}(a--f). It is observed that, compared to the Gaussian filtering method, the box filtering method allows for retaining relatively large values of $\widetilde{N_c}$ and $\widetilde{\alpha}(\nabla \widetilde{c})^2$. The results in Fig.~\ref{Fig:NCPDF}(e) and (f) show that $\widetilde{\epsilon_c}$ features statistically significant probability of both positive and negative values. As can be seen, $\widetilde{\epsilon_c}$ becomes skewed towards positive values with increasing the filter size. The positive and negative values of $\widetilde{\epsilon_c}$ as well as the impact of the filter size on these values agree with the observations reported in past numerical investigations, see for example~\cite{kasten2022modeling}. The extent of $\widetilde{\epsilon_c}$ variation and its skewness (see, Figs.~\ref{Fig:NCPDF}(e) and (f)) are of significance and suggest the importance of modeling the unresolved term for the LES of the examined turbulent premixed flames.

\begin{figure}[h!]
\centerline{\includegraphics[width=0.85\textwidth]{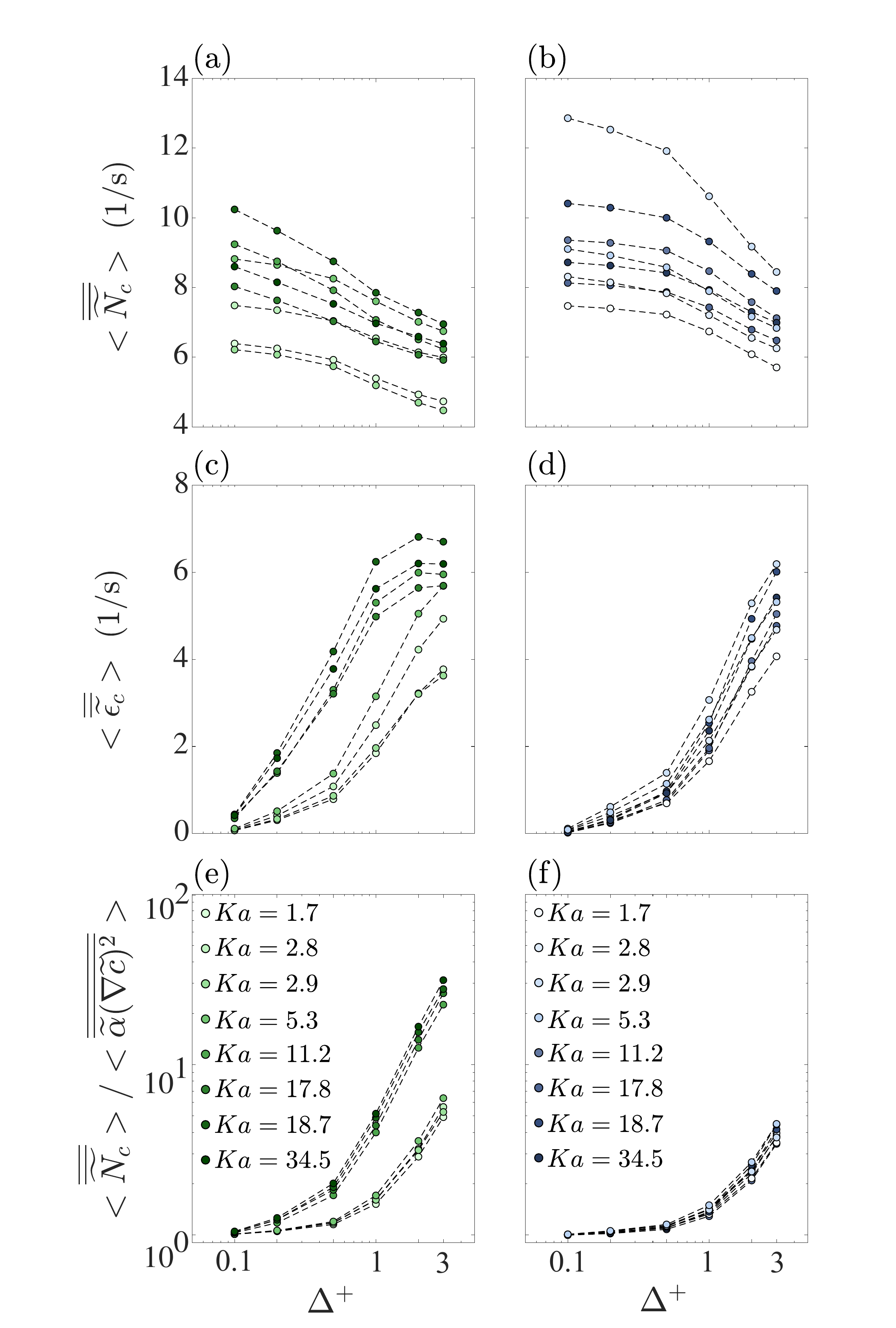}}
\caption{(a) and (b), (c) and (d), and (e) and (f) are the variations of $<\overline{\overline{\widetilde{N_c}}}>$,  $<\overline{\overline{\widetilde{\epsilon_c}}}>$, and $<\overline{\overline{\widetilde{N_c}}}>/<\overline{\overline{\widetilde{\alpha}(\nabla \widetilde{c})^2}}>$ versus the normalized filter size, respectively. The left and right columns present the results for Gaussian and box filtering methods, respectively. The symbols $<>$ and over double-bar represent the spatial and temporal averaging, respectively.}
\label{Fig:NCepsilonratio}
\end{figure}

While the above PDF analysis allows for comparing the effects of the filter type and size on $\widetilde{N_c}$, $\widetilde{\alpha}(\nabla \widetilde{c})^2$, and $\widetilde{\epsilon_c}$, the PDFs are influenced by the relatively large number of locations in the ROI corresponding to cold reactants and combustion products (which feature $\widetilde{N_c} \approx \widetilde{\alpha}(\nabla \widetilde{c})^2 \approx \widetilde{\epsilon_c} \approx0$), and this causes the most probable value of $\widetilde{N_c}$, $\widetilde{\alpha}(\nabla \widetilde{c})^2$, and $\widetilde{\epsilon_c}$ to become zero, as seen in Fig.~\ref{Fig:NCPDF}. Such bias could influence the spatially averaged values of the filtered scalar dissipation rate as well as the corresponding resolved and unresolved terms. Thus, to avoid the above bias in the averaging process, data for a small threshold ($0 \leq \widetilde{c} \leq 0.04$ and $0.96 \leq \widetilde{c} \leq 1$) were removed. The variations of $<\overline{\overline{\widetilde{N_c}}}>$ and $<\overline{\overline{\widetilde{\epsilon_c}}}>$ (with the symbol $<>$ presenting the spatial averaging and the over, double-bar symbol denoting the temporal averaging over 520 snapshots) versus the examined normalized filter size are presented in Figs.~\ref{Fig:NCepsilonratio}(a and b) and Figs.~\ref{Fig:NCepsilonratio}(c and d). The results on the left and right columns correspond to those obtained using the Gaussian and box filtering methods, respectively. It is observed that, for a given test condition, increasing $\Delta^+$ decreases $<\overline{\overline{\widetilde{N_c}}}>$. Comparison of the results in Figs.~\ref{Fig:NCepsilonratio}(a) and (b) shows that, for matching normalized filter sizes and $Ka$ numbers, $<\overline{\overline{\widetilde{N_c}}}>$ obtained from the Gaussian filtering method is smaller than that obtained from the box filtering method. The results in Figs.~\ref{Fig:NCepsilonratio}(c) and (d) show that, for both filtering methods, $<\overline{\overline{\widetilde{\epsilon_c}}}>$ generally increases with increasing $\Delta^+$, which is consistent with the pronounced skewness of the PFDs to positive values at large normalized filter sizes shown in Figs.~\ref{Fig:NCPDF}(e) and (f).

Investigating the effects of the normalized filter size and the examined test conditions on the closeness of the Favre-filtered scalar dissipation rate and the resolved term, the values of $<\overline{\overline{\widetilde{N_c}}}>/<\overline{\overline{\widetilde{\alpha}(\nabla \widetilde{c})^2}}>$ were obtained and presented in Figs.~\ref{Fig:NCepsilonratio}(e) and (f) for the Gaussian and box filtering methods, respectively. It can be seen that, $<\overline{\overline{\widetilde{N_c}}}>/<\overline{\overline{\widetilde{\alpha}(\nabla \widetilde{c})^2}}>$ obtained from the box filtering method is nearly independent of $Ka$ compared to the results obtained from the Gaussian filtering method. It is also observed that $<\overline{\overline{\widetilde{N_c}}}>/<\overline{\overline{\widetilde{\alpha}(\nabla \widetilde{c})^2}}>$ is most pronounced at the largest examined $Ka$ and for $\Delta^+ = 3.0$, with $<\overline{\overline{\widetilde{N_c}}}>/<\overline{\overline{\widetilde{\alpha}(\nabla \widetilde{c})^2}}>$ exceeding 30 and 4 for the results obtained using the Gaussian and box filtering methods, respectively. Such large values of $<\overline{\overline{\widetilde{N_c}}}>/<\overline{\overline{\widetilde{\alpha}(\nabla \widetilde{c})^2}}>$ at the large normalized filter sizes suggest the values of $<\overline{\overline{\widetilde{\epsilon_c}}}>$ approach $<\overline{\overline{\widetilde{N_c}}}>$, see Eq.~(\ref{Eq:Ncterms}). The above reported effect of $\Delta^+$ on $<\overline{\overline{\widetilde{N_c}}}>/<\overline{\overline{\widetilde{\alpha}(\nabla \widetilde{c})^2}}>$ is similar to that presented in~\cite{langella2017large,gao2014algebraic,ma2014validation,dunstan2013scalar} but extends the findings of these numerical investigations to those obtained from the experiments and to relatively larger Karlovitz numbers.

\subsection{Assessment of Eq.~(\ref{Eq:epsilon}) for modeling $\widetilde{\epsilon_c}$}\label{subsec:algebraic_unSDR}

\begin{figure}[h]
\centerline{\includegraphics[width=0.8\textwidth]{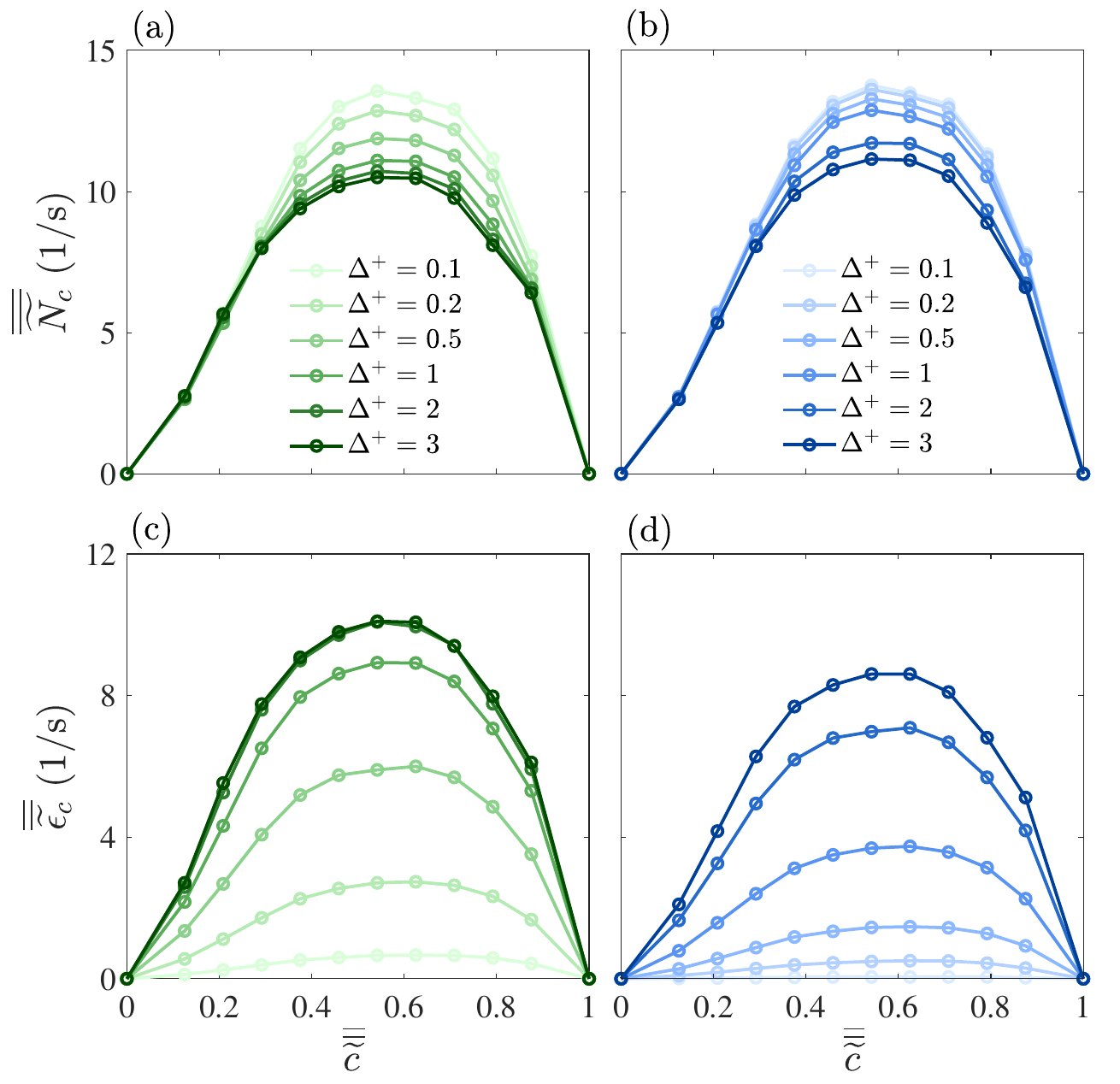}}
\caption{Conditionally averaged (a) and (b) $\overline{\overline{\widetilde{N_c}}}$ and (c) and (d) $\overline{\overline{\widetilde{\epsilon_c}}}$ versus $\overline{\overline{\widetilde{c}}}$ for Gaussian filtering (left column, in shades of green) and box filtering (right column, in shades of blue) methods and for the test condition of U7F5H40AT ($Ka = 34.5$). The normalized filter size ranges from $\Delta^+ = 0.1$ (light color) to $\Delta^+ = 3.0$ (dark color).} \label{Fig:CSDR}
\end{figure}

The temporally averaged and filtered scalar dissipation rate versus $\overline{\overline{\widetilde{c}}}$ are presented in Figs.~\ref{Fig:CSDR}(a and b) and (c and d) for the resolved and unresolved terms, respectively. The results in the left and right columns correspond to those obtained using the Gaussian and box filtering methods, respectively. As can be seen, the variations of both $\overline{\overline{\widetilde{N_c}}}$ and $\overline{\overline{\widetilde{\epsilon_c}}}$ versus $\overline{\overline{\widetilde{c}}}$ follow a parabolic-like distribution. The results in the figure show that the box filtering method yields slightly larger (smaller) values of $\overline{\overline{\widetilde{N_c}}}$ ($\overline{\overline{\widetilde{\epsilon_c}}}$) than the Gaussian filtering method at a given $\Delta^+$. Both $\overline{\overline{\widetilde{N_c}}}$ and $\overline{\overline{\widetilde{\epsilon_c}}}$ attain their maxima at $0.5 \lesssim \overline{\overline{\widetilde{c}}} \lesssim 0.6$, which is nearly independent of the examined filter size and filtering method.

\begin{figure}[h!]
\centerline{\includegraphics[width=0.75\textwidth]{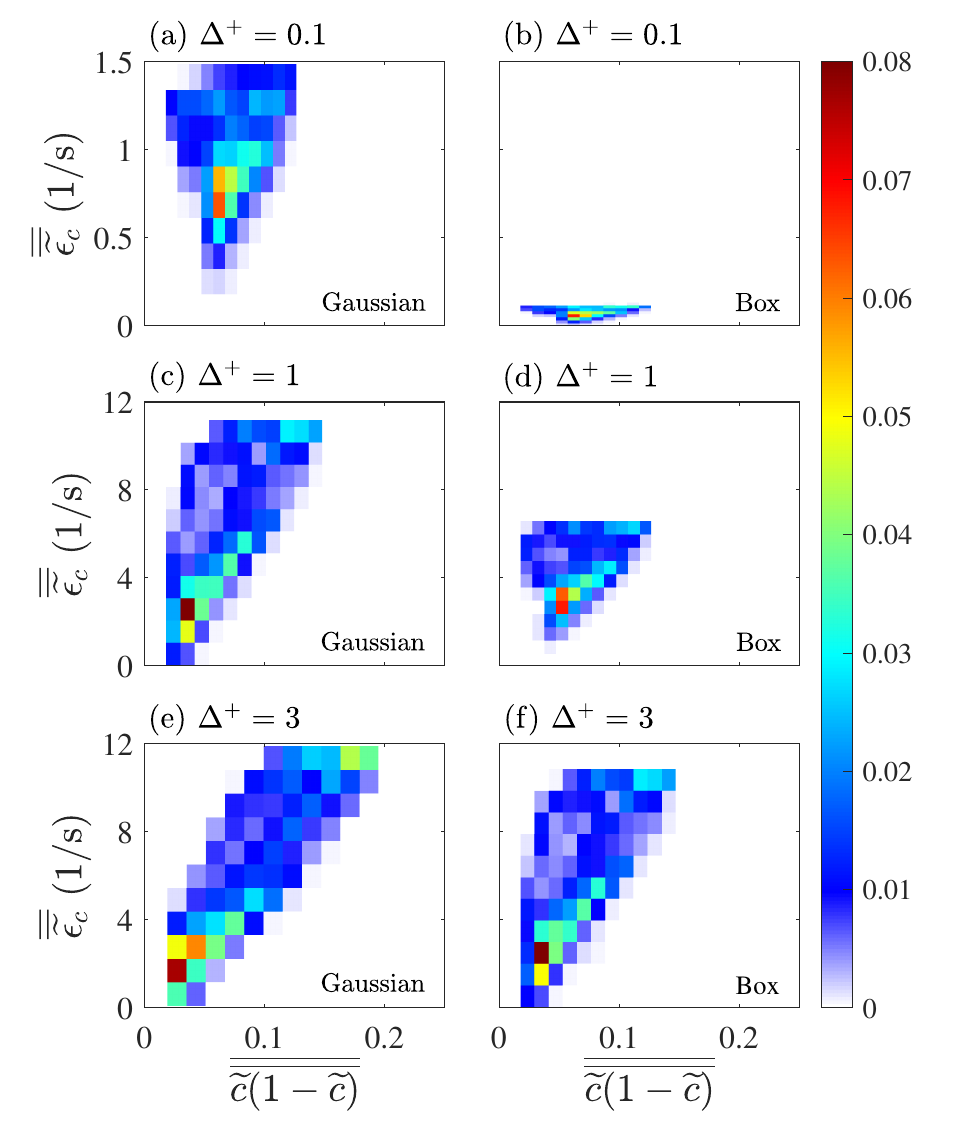}}
\caption{JPDFs of $\overline{\overline{\widetilde{\epsilon_c}}}$ and $\overline{\overline{\widetilde{c}(1-\widetilde{c})}}$ for $Ka = 34.5$. The results obtained using the Gaussian and box filtering methods are shown on the left and right columns, respectively. First, second and third rows correspond to the normalized filter sizes of 0.1, 1.0, and 3.0, respectively.}\label{Fig:JPDF}
\end{figure}

Motivated by the parabolic-like distribution of $\overline{\overline{\widetilde{\epsilon_c}}}$ versus $\overline{\overline{\widetilde{c}}}$ (as seen in Fig.~\ref{Fig:CSDR}), the Joint Probability Density Function (JPDF) of $\overline{\overline{\widetilde{\epsilon_c}}}$ and $\overline{\overline{\widetilde{c}(1-\widetilde{c})}}$ were obtained. For the representative test condition of U7F5H40AT, the JPDFs are shown in Fig.~\ref{Fig:JPDF}(a--f) for $\Delta^+ = 0.1$, 1.0, and 3.0. Similar results are obtained for the rest of the test conditions and normalized filter sizes but are not shown here for brevity. The results on the left and right columns correspond to those obtained using the Gaussian and box filtering methods, respectively. Our results show that $\overline{\overline{\widetilde{\epsilon_c}}}$ and $\overline{\overline{\widetilde{c}(1-\widetilde{c})}}$ nearly follow a linear relation, which suggests the suitability of Eq.~(\ref{Eq:epsilon}) in modeling the unresolved scalar dissipation rate for the the examined conditions of the present study. Nonetheless, it is acknowledged that deviations from the linear relation between $\overline{\overline{\widetilde{\epsilon_c}}}$ and $\overline{\overline{\widetilde{c}(1-\widetilde{c})}}$ exist, and these are anticipated to influence the accuracy of Eq.~(\ref{Eq:epsilon}) in modeling the unresolved term. This is further discussed and quantified in the following.

\begin{figure}[h]
\centerline{\includegraphics[width=0.9\textwidth]{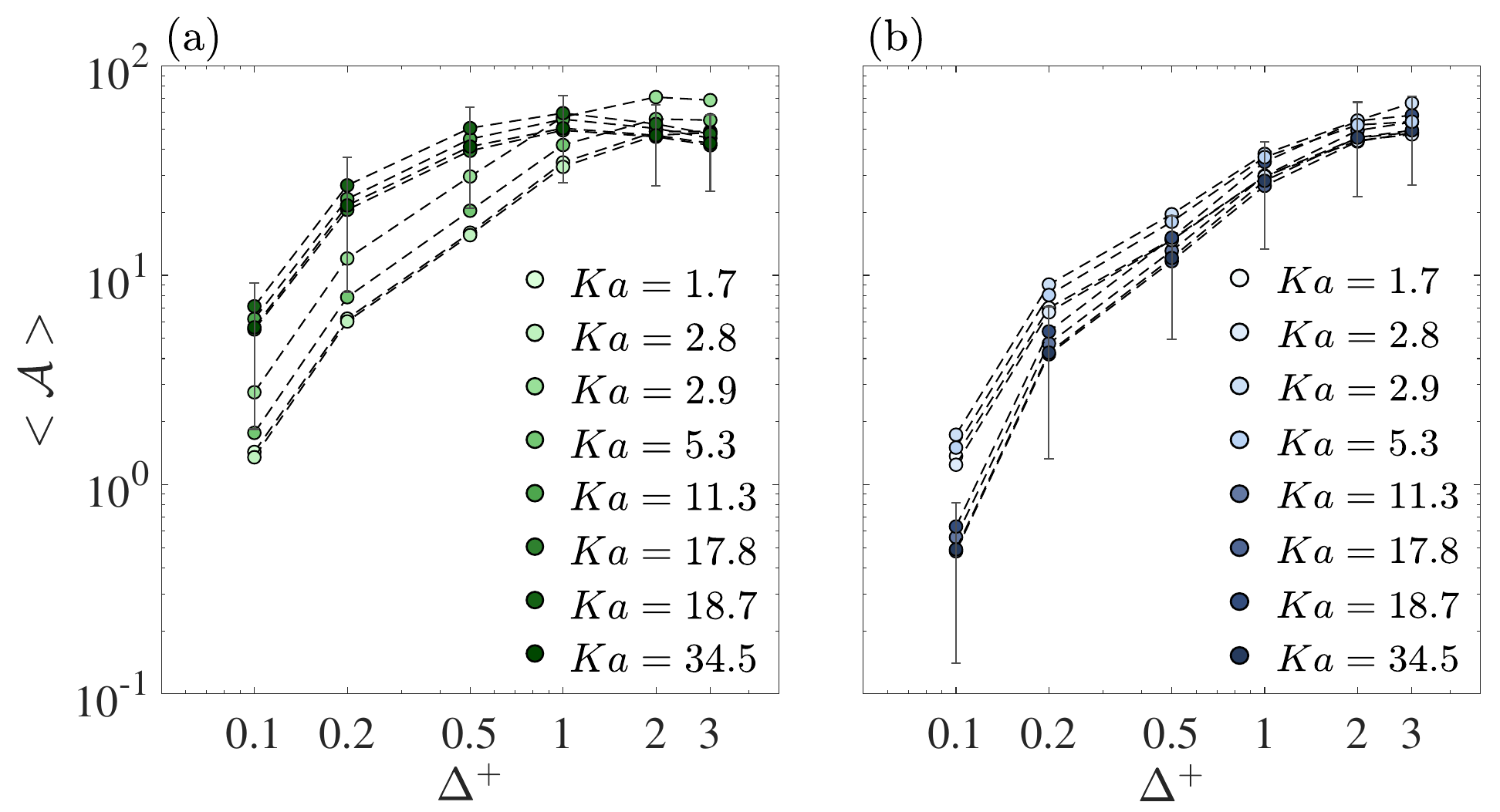}}
\caption{Variation of $<\mathcal{A}>$=$<\overline{\overline{\widetilde{\epsilon_c}}}/[\overline{\overline{\widetilde{c}(1-\widetilde{c})}}]>$ versus the normalized LES filter sizes for all examined test conditions, using (a) Gaussian and (b) box filtering methods.}
\label{Fig:A}
\end{figure}

For all examined test conditions and normalized LES filter sizes, the values of $\mathcal{A}=\overline{\overline{\widetilde{\epsilon_c}}}/[\overline{\overline{\widetilde{c}(1-\widetilde{c})}}]$ were calculated; and, the spatially averaged values of $\mathcal{A}$ were presented in Figs.~\ref{Fig:A}(a) and (b) for the Gaussian and box filtering methods, respectively. At a fixed normalized filter size and for all examined test conditions, the standard deviation of $\mathcal{A}$ was calculated, and the maximum value for the corresponding $Ka$ was presented using the error bars in Fig.~\ref{Fig:A}. For both utilized LES filter types, the largest error bar corresponds to the test condition with $Ka = 34.5$ and for all filter sizes, except for the normalized filter sizes of 0.1 and 2.0, that the largest standard deviation of $\mathcal{A}$ corresponded to $Ka = 17.8$. For $Ka \leq 5.3$, the results in Fig.~\ref{Fig:A}(a) show that increasing $\Delta^+$ increases $<\mathcal{A}>$ obtained from the Gaussian filtering method. For this filtering method and for $Ka > 5.3$, increasing the normalized filter size from 0.1 to 1.0 increases $<\mathcal{A}>$; however, this parameter decreases with increasing the normalized filter size beyond unity. Compared to the results obtained for the Gaussian-filtered data, the results obtained from the box filtering method show that $<\mathcal{A}>$ increases with increasing $\Delta^+$ and for all examined test conditions. It is observed that, compared to $<\mathcal{A}>$ obtained from the Gaussian filtering method, that obtained from the box filtering method features a smaller sensitivity to the examined test condition at a fixed normalized filter size. This is a desirable characteristic for the LES of the turbulent premixed flames. This finding may suggest, despite the model in Eq.~(\ref{Eq:epsilon}) was originally developed and examined for Gaussian filtered parameters, robust predictions may also be obtained using the box filtering method.

\begin{figure}[h]
\centerline{\includegraphics[width=1\textwidth]{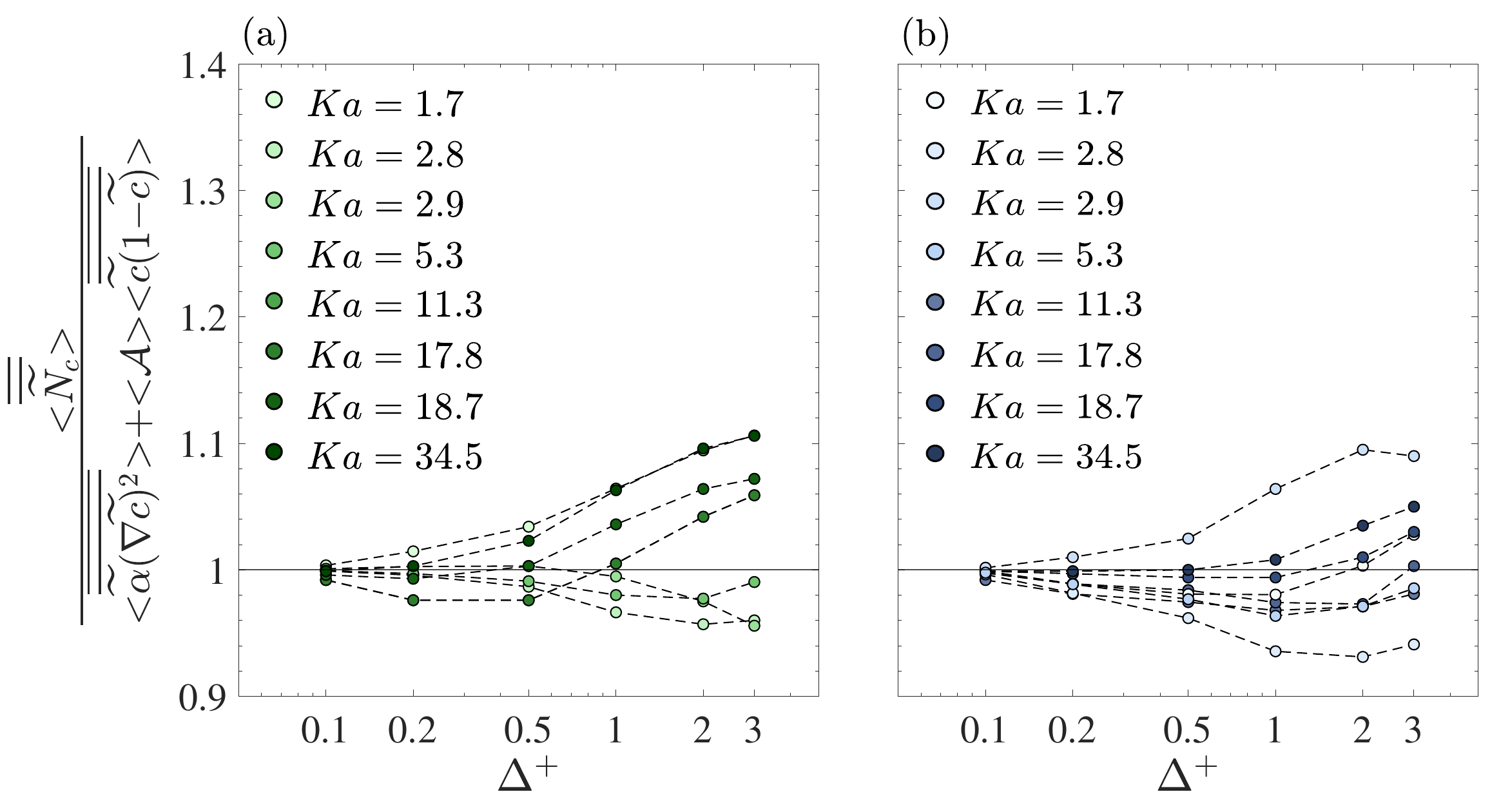}}
\caption{Ratio of $<\overline{\overline{\widetilde{N_c}}}>$ and $<\overline{\overline{\widetilde{\alpha}(\nabla\widetilde{c})^2}}>+<\mathcal{A}><\overline{\overline{\widetilde{c}(1-\widetilde{c})}}>$ for (a) Gaussian and (b) box filtering methods. The symbols $<>$ and over double-bar represent the spatial and temporal averaging, respectively.}\label{Fig:ratio}
\end{figure}

For a given test condition, it may be anticipated that the argument in the curly braces of Eq.~(\ref{Eq:epsilon}) should be independent of space and time. As such, after temporally and spatially averaging both sides of Eqs.~(\ref{Eq:Ncterms}) and (\ref{Eq:epsilon}), it would be expected that $<\overline{\overline{\widetilde{N_c}}}>$ equates $<\overline{\overline{\widetilde{\alpha}(\nabla \widetilde{c})^2}}>+<\mathcal{A}>< \overline{\overline{\widetilde{c}(1-\widetilde{c})}}>$. However, differences between these two parameters exist as a result of spatial variations in $\mathcal{A}$, with the characteristics of such variation of $\mathcal{A}$ discussed and quantified in Appendix~A. It is of interest to assess if the estimated values of $<\mathcal{A}>$ would facilitate acceptable prediction of the unresolved term. To this end, the ratios of $<\overline{\overline{\widetilde{N_c}}}>$ and $<\overline{\overline{\widetilde{\alpha}(\nabla\widetilde{c})^2}}>+<\mathcal{A}><\overline{\overline{\widetilde{c}(1-\widetilde{c})}}>$ were calculated, with these ratios presented in Fig.~\ref{Fig:ratio}(a) and (b) for the Gaussian and box filtering methods, respectively. In the figure, close-to-unity values of $<\overline{\overline{\widetilde{N_c}}}>/[<\overline{\overline{\widetilde{\alpha}(\nabla\widetilde{c})^2}}>+<\mathcal{A}><\overline{\overline{\widetilde{c}(1-\widetilde{c})}}>]$ would suggest that the algebraic model provides an acceptable estimation of the unresolved term and as a result $<\overline{\overline{\widetilde{N_c}}}>$. As can be seen in Fig.~\ref{Fig:ratio}, the predictions of the model in Eq.~(\ref{Eq:epsilon}) could deviate from $<\overline{\overline{\widetilde{N_c}}}>$ (i.e. pronounced deviation of the ratio from unity) with increasing $\Delta^+$ for both filtering methods. The results in Fig.~\ref{Fig:ratio} show that the maximum reported deviation of $<\overline{\overline{\widetilde{N_c}}}>/[<\overline{\overline{\widetilde{\alpha}(\nabla\widetilde{c})^2}}>+<\mathcal{A}><\overline{\overline{\widetilde{c}(1-\widetilde{c})}}>]$ from unity is 11\%, with the majority of the examined test conditions and normalized filter sizes featuring a deviation less than 5\%. The results shown in Figs.~\ref{Fig:ratio} and \ref{Fig:NCepsilonratio} indicate that, although different filtering methods yield different values of $<\overline{\overline{\widetilde{N_c}}}>$ and $<\overline{\overline{\widetilde{N_c}}}>/<\overline{\overline{\widetilde{\alpha}(\nabla \widetilde{c})^2}}>$, the value of $<\overline{\overline{\widetilde{N_c}}}>/[<\overline{\overline{\widetilde{\alpha}(\nabla\widetilde{c})^2}}>+<\mathcal{A}><\overline{\overline{\widetilde{c}(1-\widetilde{c})}}>]$ obtained from both filtering approaches exhibit similar deviations from unity. It is concluded that the algebraic model of Eq.~(\ref{Eq:epsilon}) provides acceptable predictions of the unresolved scalar dissipation rate for the examined conditions of the present study and for both filtering methods.

The experimental results of the present study were analyzed to gain insight into the modeling terms inside the curly braces of Eq.~(\ref{Eq:epsilon}), specifically $\beta_c$. It was assumed that the argument inside the curly braces of Eq.~(\ref{Eq:epsilon}) is spatially and temporally invariant and equals $<\mathcal{A}>$. Then, the left- and right-hand-sides of Eq.~(\ref{Eq:epsilon}) were spatially and temporally averaged, yielding

\begin{equation}
\label{Eq:betac}
\beta_c =\frac{1}{<\mathcal{A}>}\left\{\underbrace{2\mathcal{F}K_c\frac{S_\mathrm{L,0}}{\delta_\mathrm{th}}}_{T_1} + \underbrace{\mathcal{F}(C_3-\tau C_4 Da_\Delta)(\frac{2u'_\Delta}{3\Delta})}_{T_2}\right\}.
\end{equation}
In Eq.~(\ref{Eq:betac}), $T_1/\mathcal{F}$ solely depends on the fuel and air chemistry; and, $T_2/\mathcal{F}$ additionally depends on the background flow turbulence and the LES filter size. As such, in addition to the experimentally estimated value of $<\mathcal{A}>$, Eq.~(\ref{Eq:betac}) suggests that
$\beta_c$ would depend on $S_\mathrm{L,0}$, $\delta_\mathrm{th}$, and $\tau$ as well as $\Delta$ and $u'_{\Delta}$. Key to estimating $\beta_c$ in Eq.~(\ref{Eq:betac}) is the subgrid scale velocity fluctuations RMS, which is estimated in the present study using the non-reacting velocity data. In an earlier investigation~\cite{mohammadnejad2022new}, we showed that the spectrum of the background turbulent kinetic energy follows the $-5/3$ decay for $ 0.01 <  2\pi \eta/l < 0.1$, with $l$ being a characteristic length scale. For such power-law decay of the turbulent kinetic energy, the SGS velocity fluctuations RMS is given by $u'_\Delta \approx (\epsilon_\mathrm{K} \Delta)^{1/3}$, with $\epsilon_\mathrm{K}$ being the dissipation rate of the turbulent kinetic energy~\cite{pope2001turbulent,ghosal1995dynamic}. For the examined conditions of the present study, Mohammadnejad \textit{et al.}~\cite{mohammadnejad2022new} showed that the normalized energy dissipation rate $C_\epsilon = \epsilon_\mathrm{K}/(u'^3/L) = \mathcal{O}(1)$, suggesting $\epsilon_\mathrm{K} \approx u'^3/L$. As a result, the subgrid scale velocity fluctuations RMS of the conditions examined in the present study could be approximated using $u'_\Delta \approx [(u'^3/L) \Delta]^{1/3} \approx u'(\Delta/L)^{1/3}$. Such a power-law relation between $u'_\Delta/u'$ and $\Delta/L$ is consistent with that used in past studies~\cite{steinberg2010stretch}. Two limitations are acknowledged that influence the estimation of $u'_\Delta$ in the present study. First, $u'_\Delta$ should be ideally obtained for reacting flow conditions. Langella \textit{et al.}~\cite{langella2017large} filtered the DNS of turbulent premixed flames to evaluate $u'_\Delta$; and, it was shown in~\cite{langella2017large} this parameter remains nearly constant within the flame brush. Second, the minimum value of $\Delta$ that could be assessed in the present study is the velocity measurements resolution, which is the length of the wire in our hotwire anemometry. As such, the values of $\Delta^+$ that could be used for evaluating $\beta_c$ are $\Delta^+> l_\mathrm{w}/\delta_\mathrm{th} = 1.25/0.66 =1.89$ and $\Delta^+> 1.25/1.29 = 0.97$ for our examined methane-air and hydrogen-enriched methane-air flames, respectively. Among the examined normalized filter sizes, $\Delta^+ = 2.0$ and 3.0 satisfy the above conditions and for both examined fuel and air compositions.

\begin{figure}[h!]
	\centerline{\includegraphics[width=1\textwidth]{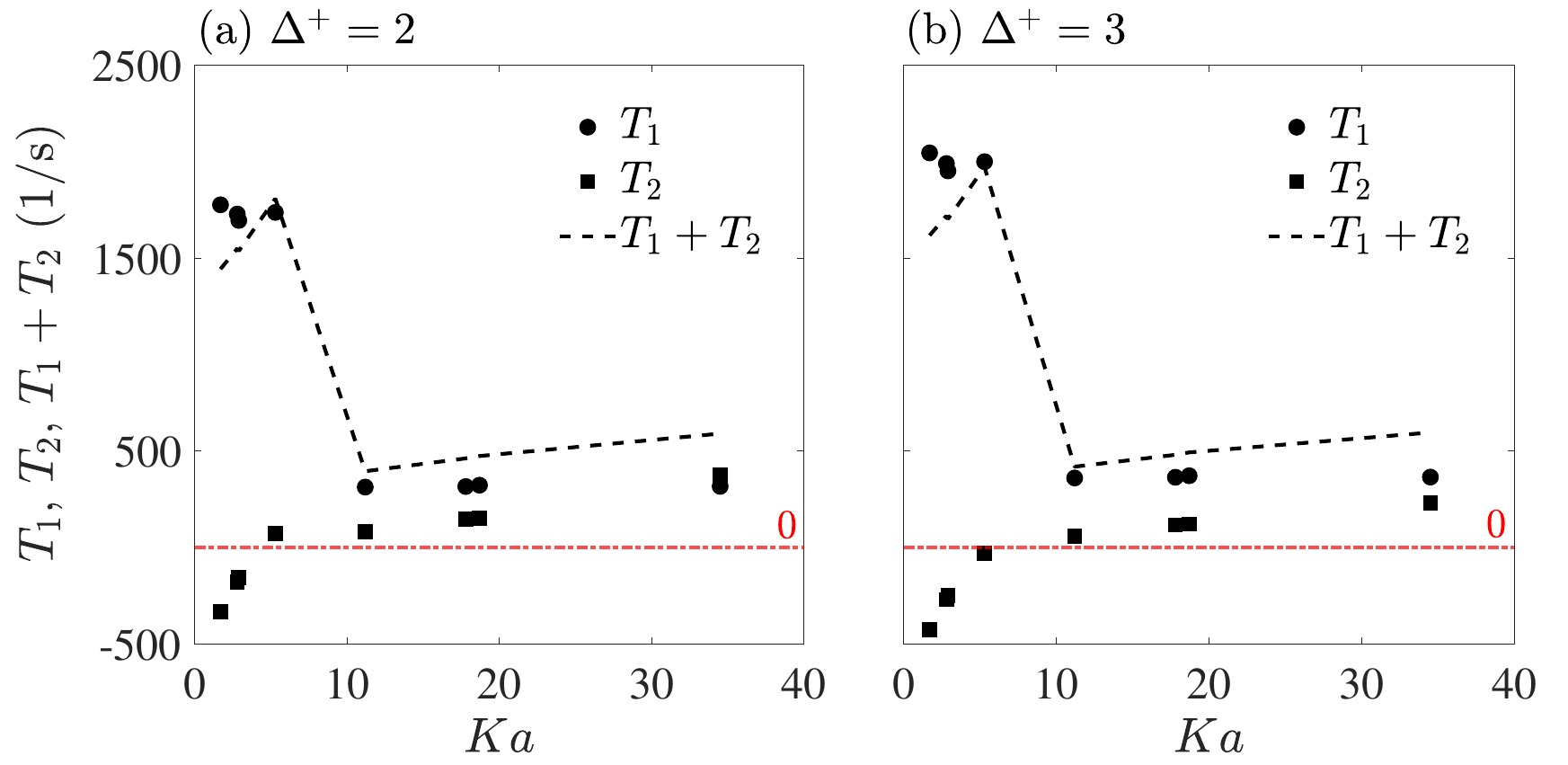}}
	\caption{Variation of $T_1$, $T_2$, and $T_1+T_2$ with $T_1=2\mathcal{F}K_c S_\mathrm{L,0}/\delta_\mathrm{th}$, and $T_2=\mathcal{F}(C_3-\tau C_4 Da_\Delta)[2u'_\Delta/(3\Delta)]$, for normalized filter sizes of (a) $\Delta^+ = 2.0$, and (b) $\Delta^+ = 3.0$. The red dotted-dashed line shows the zero value on the vertical axis.} \label{Fig:slopeterms}
\end{figure}

For the fuel and air compositions of the present study as well as $\Delta^+ = 2.0$ and 3.0, $u'_\Delta = u'(\Delta/L)^{1/3}$ was obtained and $T_1$ and $T_2$ were estimated. Figures~\ref{Fig:slopeterms}(a) and (b) present the variations $T_1$, $T_2$, and $T_1+T_2$ versus the examined $Ka$, using the circular data symbol, square data symbol, and black dashed line, respectively. The red dotted-dashed line is the zero value on the vertical axis. The results in Figs.~\ref{Fig:slopeterms}(a) and (b) correspond to $\Delta^+ = 2.0$ and 3.0, respectively. It is observed that, for $1.7 \leq Ka \leq 5.3$ (which correspond to our examined methane-air flames), $T_1$ is significantly larger than those for $11.2 \leq Ka \leq 34.5$ (which correspond to the hydrogen-enriched methane-air flames). This is because the values of $S_\mathrm{L,0}/\delta_\mathrm{th}$ significantly differ between our examined methane-air (with $S_\mathrm{L,0}/\delta_\mathrm{th}$ = \unit[288]{1/s}) and hydrogen-enriched methane-air (with $S_\mathrm{L,0}/\delta_\mathrm{th}$ = \unit[62]{1/s}) flames. Such large values of $S_\mathrm{L,0}/\delta_\mathrm{th}$ for the examined methane-air flames contribute to the relatively large values of $T_1$ and as a result $T_1+T_2$. Using the definitions of $C_3$ and $C_4$, it can be shown that approaching the laminar flame condition (i.e., $u'_{\Delta} \rightarrow 0$), $C_3 \rightarrow 0$ and $C_4 \rightarrow 1.1$. As a result, while $T_1 = 2\mathcal{F}K_cS_\mathrm{L,0}/\delta_\mathrm{th}$, $T_2 \rightarrow -1.1\mathcal{F}\tau (u'_\Delta/S_\mathrm{L,0})^{-1}(\Delta/\delta_\mathrm{th})[2u'_\Delta/(3\Delta)]= -0.73\mathcal{F}\tau S_\mathrm{L,0}/\delta_\mathrm{th}$. At the limit of $u'_{\Delta} \rightarrow 0$, $T_1/T_2 \rightarrow -(2/0.73)K_c/\tau = -(2/0.73)\times0.79 = -2.2$. Thus, at this limit, $|T_2|$ is about half of $T_1$, leading to $T_1+T_2 \approx 0.5T_1$. As a result, provided $T_1$ is large, $T_1+T_2$ is rather large at the limit of $u'_{\Delta} \rightarrow 0$. Although $u'_{\Delta}$ does not equate to zero even for the smallest examined Karlovitz number flames of the present study, the negative values of $T_2$ and large values of $T_1$ (as well as $T_1+T_2$) are evident in Fig.~\ref{Fig:slopeterms} as $Ka$ decreases (see data points with $Ka<5.3$ in the figure). Compared to the examined methane-air flames of the present study, however, the examined hydrogen-enriched methane-air flames (which also correspond to larger examined Karlovitz numbers) feature relatively smaller values of $T_1+T_2$. This finding has implications for estimation of $\beta_c$.

Equation~(\ref{Eq:betac}) along with the values of $T_1$ and $T_2$ as well as the experimentally estimated values of $<\mathcal{A}>$ were used to obtain $\beta_c$. It is important to note that the values of the latter can be influenced by the 2D characteristic of our diagnostics versus the 3D nature of the examined flames. For isotropic background flow, statistically stationary flames, unity Lewis number fuel and air mixtures, and $Re_\mathrm{T}>100$, Chakraborty \textit{et al.}~\cite{chakraborty2013determination} showed that $\widetilde{N_c}$ estimated using 3D data ($\widetilde{N_{c,\mathrm{3D}}}$) relates to the corresponding values estimated in the 2D space ($\widetilde{N_{c,\mathrm{2D}}}$) via $\widetilde{N_{c,\mathrm{3D}}}\approx1.5\widetilde{N_{c,\mathrm{2D}}}$. Since this approximation is independent of $\widetilde{c}$, it can be spatially and temporally averaged, yielding $<\overline{\overline{\widetilde{N_{c,\mathrm{3D}}}}}> \approx1.5<\overline{\overline{\widetilde{N_{c,\mathrm{2D}}}}}>$. Indeed, in the present study, the effective Lewis number is about unity, $Re_\mathrm{T}>100$ (see~Table~\ref{tab:Conditions}), and the examined flames are statistically stationary. For the examined conditions of the present study, it was shown in our previous work~\cite{mohammadnejad2022new} that the exponent of $(u'/U)^2$ power-law decay versus $y$ yields values less than -1, which agree with those reported for isotropic turbulence produced by passive and active generators~\cite{lavoie2007effects,hearst2014decay}. It is noted that, for $\Delta^+=2.0$ and 3.0, the results presented in Figs.~\ref{Fig:NCepsilonratio}(e) and (f) showed that   $<\overline{\overline{\widetilde{N_c}}}>/<\overline{\overline{\widetilde{\alpha}(\nabla \widetilde{c})^2}}>=1+<\overline{\overline{\widetilde{\epsilon_c}}}>/<\overline{\overline{\widetilde{\alpha}(\nabla \widetilde{c})^2}}>$ is larger than 2 for both Gaussian and box filtering methods. This ratio further increases by approximately one order of magnitude for $\Delta^+\geq2.0$ and $Ka\geq11.2$ for Gaussian filtered results, indicating that $<\overline{\overline{\widetilde{\epsilon_c}}}>$ becomes significantly larger than $<\overline{\overline{\widetilde{\alpha}(\nabla \widetilde{c})^2}}>$ for these conditions. This along with the definition of the unresolved scalar dissipation rate (see, Eq.~(\ref{Eq:Ncterms})) would suggest that $<\overline{\overline{\widetilde{\epsilon_{c,\mathrm{3D}}}}}>\approx1.5<\overline{\overline{\widetilde{\epsilon_{c,\mathrm{2D}}}}}>$ at large normalized filter sizes. Compared to the effect of 2D nature of the measurements on evaluating the SGS scalar dissipation rate, such nature is not anticipated to influence the Favre-averaged progress variable fields and as a result $<\overline{\overline{\widetilde{c}(1-\widetilde{c})}}>$. As such, it is concluded that the values of $<\mathcal{A}>$ should be scaled by a factor of 1.5, correcting for the 2D nature of the measurements. That is, in Eq.~(\ref{Eq:betac}), the values of $\beta_c$ in 3D ($\beta_{c,\mathrm{3D}}$) will have to be multiplied by $1/1.5 = 2/3$ provided $<\mathcal{A}>$ is calculated form the 2D measurements.

\begin{figure}[h]
	\centerline{\includegraphics[width=1\textwidth]{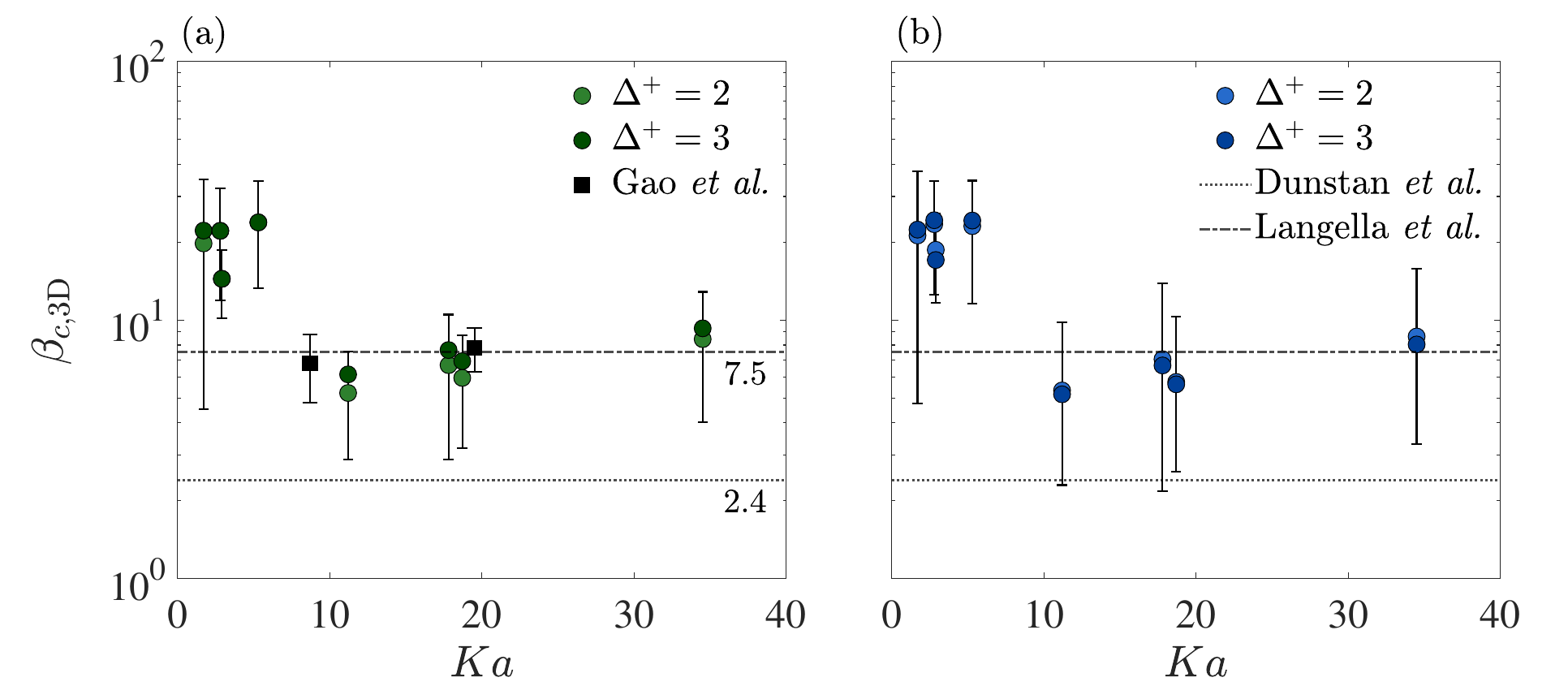}}
	\caption{Values of $\beta_{c,\mathrm{3D}}$ estimated using the experimentally obtained values of $< \mathcal{A} >$ for (a) Gaussian and (b) box filtering methods versus $Ka$, for $\Delta^+=2$ and 3. The square data points in panel (a) are reported for $Le=1$, $Ka=8.7$ and $19.5$, and $\Delta^+=2.8$ in~\cite{gao2015dynamic}. The horizontal dotted and dotted-dashed lines at $\beta_{c,\mathrm{3D}}=2.4$ and $7.5$ are the values in~\cite{dunstan2013scalar} and~\cite{langella2017large}, respectively.} \label{Fig:comparison}
\end{figure}

Figures~\ref{Fig:comparison}(a) and (b) present the values of $\beta_{c,\mathrm{3D}}$ for the results filtered using the Gaussian and box filtering methods, respectively. In the figure, the extent of the error bars reflects the effect of $\mathcal{A}$ spatial variation on $\beta_{c,\mathrm{3D}}$ at a given $Ka$. For each $Ka$, the largest error-bar is shown in Fig.~\ref{Fig:comparison}. The largest error bars corresponded to $\Delta^+=2.0$ in the present study. Also, overlaid on Fig.~\ref{Fig:comparison}(a) are the square data points, which are the spatially averaged values of $\beta_{c,\mathrm{3D}}$ reported in Gao \textit{et al.}~\cite{gao2015dynamic}. Specifically, for a Gaussian filter, unity Lewis number flames, and $\Delta^+=2.8$, the spatially averaged values of $\beta_{c,\mathrm{3D}}$ are 6.7 and 7.8 for $Ka=8.7$ and $19.5$, respectively, in~\cite{gao2015dynamic}. The error bars for these values of $\beta_{c,\mathrm{3D}}$ (overlaid on the square data points) are extracted form~\cite{gao2015dynamic}. In Fig.~\ref{Fig:comparison}, the horizontal dotted and dotted-dashed lines are $\beta_{c,\mathrm{3D}}=2.4$ and $7.5$, which are those reported in the numerical studies of Dunstan \textit{et al.}~\cite{dunstan2013scalar} and Langella \textit{et al.}~\cite{langella2017large}, respectively.

Figures~\ref{Fig:comparison}(a) and (b) show that, for $Ka \leq 5.3$ (which correspond to the examined 
methane-air flames), $\beta_{c,\mathrm{3D}}$ varies between 14 and 24, which are rather large compared to those reported in~\cite{gao2015dynamic,langella2017large,dunstan2013scalar}. The large values of $T_1+T_2$ (see, Fig.~\ref{Fig:slopeterms}) combined with the nearly constant experimentally obtained values of $<\mathcal{A}>$ at the relatively large examined filter sizes (see, Fig.~\ref{Fig:A}) result in the large values of $\beta_{c,\mathrm{3D}}$ for the examined small Karlovitz number flames. Compared to these, however, for $Ka \geq 11.2$ (which correspond to the examined hydrogen-enriched methane-air flames), the results in Figs.~\ref{Fig:comparison}(a) and (b) show that the experimentally obtained values of $\beta_{c,\mathrm{3D}}$ slightly increase from about 5 to 9 with increasing $Ka$. Given the extent of the error bars, for $Ka \geq 11.2$, the obtained values of $\beta_{c,\mathrm{3D}}$ agree well with those reported in the literature~\cite{gao2015dynamic,langella2017large}. Despite the experimental limitations of our study in measuring the SGS scalar dissipation rate as well as the differences between the flame configuration of our investigation (V-shaped flames) and those of past numerical investigations (flame-in-a-box~\cite{gao2015dynamic} and jet flame~\cite{langella2017large}), the similarity between $\beta_{c,\mathrm{3D}}$ values at matching Karlovitz numbers are notable. This may suggest that the model proposed in Eq.~(\ref{Eq:epsilon}) is flame-configuration independent and suitable for predicting the SGS scalar dissipation rate at matching background flow and fuel conditions. Future investigations will aim to address diagnostics limitations and extending the examined test conditions.

\section{Summary and conclusions}\label{sec 5}

A sub-grid-scale closure for the scalar dissipation rate ($N_c$) modeling used in the Large Eddy Simulations of turbulent premixed flames was studied experimentally. This closure was originally proposed by \cite{kolla2009scalar} in the framework of Favre-averaged Navier Stokes and later extended to the LES framework by \cite{dunstan2013scalar}. Our experiments were performed using the hotwire anemometry technique to characterize the background turbulent flow, and the Rayleigh scattering to measure the scalar dissipation rate. The fuel was either methane or a mixture of 60\% (by volume) methane and 40\% hydrogen. The Karlovitz number varied between 1.7 and 34.5. The experimentally obtained scalar dissipation rate fields were filtered using LES filter sizes that varied between 0.1 to 3.0 times the thermal thickness of the examined laminar flames as well as using Gaussian and box filtering methods. 

The Probability Density Function (PDF) of the Favre-filtered scalar dissipation rate ($\widetilde{N_c}$) as well as those of the resolved ($\widetilde{\alpha}(\nabla\widetilde{c})^2$) and unresolved ($\widetilde{\epsilon_c}$) scalar dissipation rate  became generally broader by increasing the LES filter size. The PDF of $\widetilde{\epsilon_c}$ was asymmetric and skewed towards positive values. Consistent with the above finding, the ratio of temporally and spatially averaged filtered scalar dissipation rate and its resolved term ($<\overline{\overline{\widetilde{N_c}}}>/<\overline{\overline{\widetilde{\alpha}(\nabla\widetilde{c})^2}}>$) increased to several folds with increasing the normalized filter size. This increase was more pronounced for the results obtained using the Gaussian filtering method. These observations suggested the importance of modeling the unresolved term for predicting $\widetilde{N_c}$. 

The variation of the temporally averaged scalar dissipation rate ($\overline{\overline{\widetilde{\epsilon_c}}}$) versus the temporally averaged combustion progress variable ($\overline{\overline{\widetilde{c}}}$) 
followed a parabolic-like distribution. Motivated by this, the Joint Probability Density Function analysis was performed and it was shown $\overline{\overline{\widetilde{\epsilon_c}}}$ follows a nearly linear relation with $\overline{\overline{\widetilde{c}(1-\widetilde{c})}}$. The slopes of such relations were obtained and used to estimate the temporally and spatially averaged unresolved term. The summation of the resolved and modeled unresolved terms was calculated, and compared with the measured filtered scalar dissipation rate. It was obtained that the algebraic model yielded acceptable prediction of the filtered scalar dissipation rate with a maximum deviation from the measured values of about $11\%$. This deviation was generally more pronounced at the larger examined normalized filter sizes. This conclusion was consistent for both examined filtering methods.

Our experimental measurements were used to estimate a key parameter, $\beta_{c,\mathrm{3D}}$, in the algebraic model used for predicting the unresolved scalar dissipation rate. While, $\beta_{c,\mathrm{3D}}$ obtained from the experimental measurements was relatively large for the examined small Karlovitz number flames, $\beta_{c,\mathrm{3D}}$ was similar to those reported in past numerical investigations and for the examined flames with relatively large Karlovitz numbers. Specifically, increasing the Karlovitz number from 11.2 to 34.5 increased $\beta_{c,\mathrm{3D}}$ from about 5 to 9, which is similar to 7.5 reported in \cite{langella2017large}. This finding was shown to be independent of the examined filter type. To our best knowledge, the present study experimentally demonstrated for the first time that the measured values of $\beta_{c,\mathrm{3D}}$ are similar to those obtained from the filtered DNS for other flame configurations and are independent of the examined filtering method. While the results of the present study experimentally demonstrated the suitability of the algebraic sub-grid-scale closure for modeling the scalar dissipation rate, our findings are limited to the capabilities of our measurement techniques and extent of the examined test conditions. Future work will aim at addressing these limitations.


\section*{Authors contributions}
\textbf{Morteza Nahvi}: Writing-review \& editing, Writing-original draft, Visualization, Methodology, Investigation, Formal analysis, Data curation. \textbf{Sina Kheirkhah}: Writing-review \& editing,
Writing-original draft, Supervision, Resources, Project administration,
Methodology, Investigation, Funding acquisition.

\section*{Acknowledgments}

The authors are thankful for the financial support from Alberta Innovates--Hydrogen Center of Excellence (HCOE2), Natural Sciences and Engineering Research Council of Canada (RGPIN-2025-04887), and Canada Foundation for Innovation (37470).

\appendix
\section*{Appendix A: Effect of the spatial variation of $\mathcal{A}$ on predicting the filtered scalar dissipation rate}
\label{app:spatial_variation}

\renewcommand{\thefigure}{A.\arabic{figure}}
\setcounter{figure}{0}

\begin{figure}[h!]
\centerline{\includegraphics[width=0.8\textwidth]{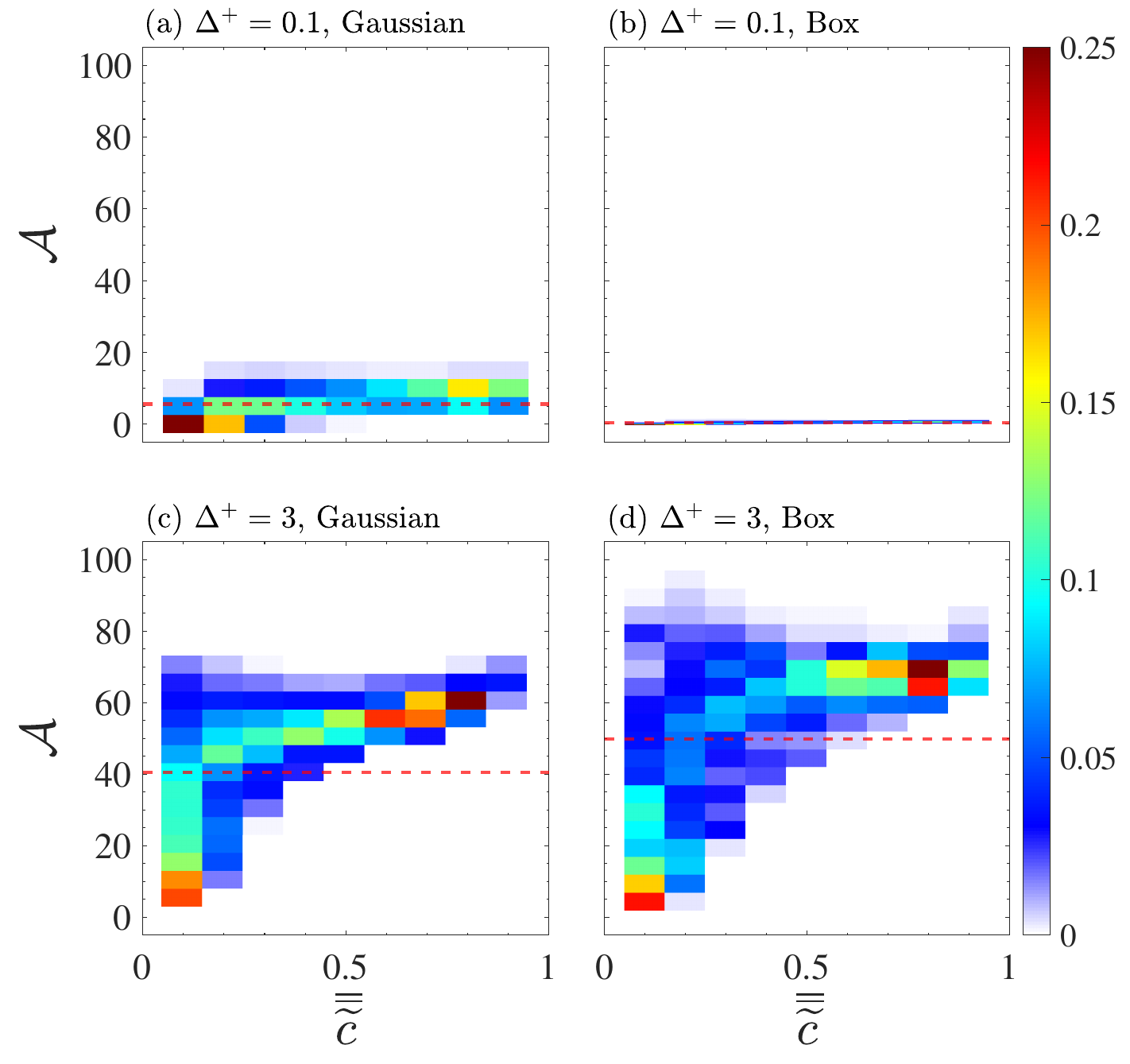}}
\caption{JPDFs of $\mathcal{A}$ and $\overline{\overline{\widetilde{c}}}$ for $Ka=34.5$. (a) and (b) are for $\Delta^+=0.1$ and (c) and (d) are for $\Delta^+=3.0$. Left and right columns correspond to results obtained from the Gaussian and box filtering methods, respectively. The red dashed line represents the spatially averaged value of $\mathcal{A}$.}
\label{Fig:Aspace}
\end{figure}

Although the results of the present study suggest that the spatially averaged value of $\mathcal{A}$ would yield acceptable prediction of $<\overline{\overline{\widetilde{N_c}}}>$ (see, Fig.~\ref{Fig:ratio}), the sensitivity of predicting this parameter to the spatial variation of $\mathcal{A}$ should be investigated. To this end, JPDFs of $\mathcal{A}$ and $\overline{\overline{\widetilde{c}}}$ were examined for several test conditions. Figure~\ref{Fig:Aspace} presents the JPDFs of $\mathcal{A}$ and $\overline{\overline{\widetilde{c}}}$ for $Ka=34.5$, $\Delta^+=0.1$ and $3.0$, and for both the Gaussian and box filtering methods. In the figure, the red dashed line presents the spatially averaged value of $\mathcal{A}$. It should be noted that similar JPDFs were obtained for the other test conditions and the results for $Ka = 34.5$ are presented here for brevity. Comparison of the results in the first and second rows of Fig.~\ref{Fig:Aspace} shows that, for both filtering methods, increasing the normalized filter size increases the values of $\mathcal{A}$ (and as a result $<\mathcal{A}>$), which is consistent with the results in Fig.~\ref{Fig:A}. Figure~\ref{Fig:Aspace} shows that increasing $\Delta^+$ increases the scatter of $\mathcal{A}$, especially closer to the fresh gas, with $\overline{\overline{\widetilde{c}}} \lesssim 0.4$. Comparing the results in Figs.~\ref{Fig:Aspace}(c) and (d) shows that $\mathcal{A}$ exhibits larger scatter for the box filtering method (compared to Gaussian filtering method) and for $\Delta^+=3.0$. For both filtering methods, values of $\mathcal{A}$ with large probabilities near $\overline{\overline{\widetilde{c}}} \gtrsim 0.7$ are observed. The results in Fig.~\ref{Fig:Aspace} show that the deviation of $\mathcal{A}$ from its spatially averaged value is less than 20\% (35\%) for $\Delta^+=0.1$ ($\Delta^+=3.0$) and for $0.5 \lesssim \overline{\overline{\widetilde{c}}}\lesssim 0.9$.

\begin{figure}[h!]
\centerline{\includegraphics[width=0.85\textwidth]{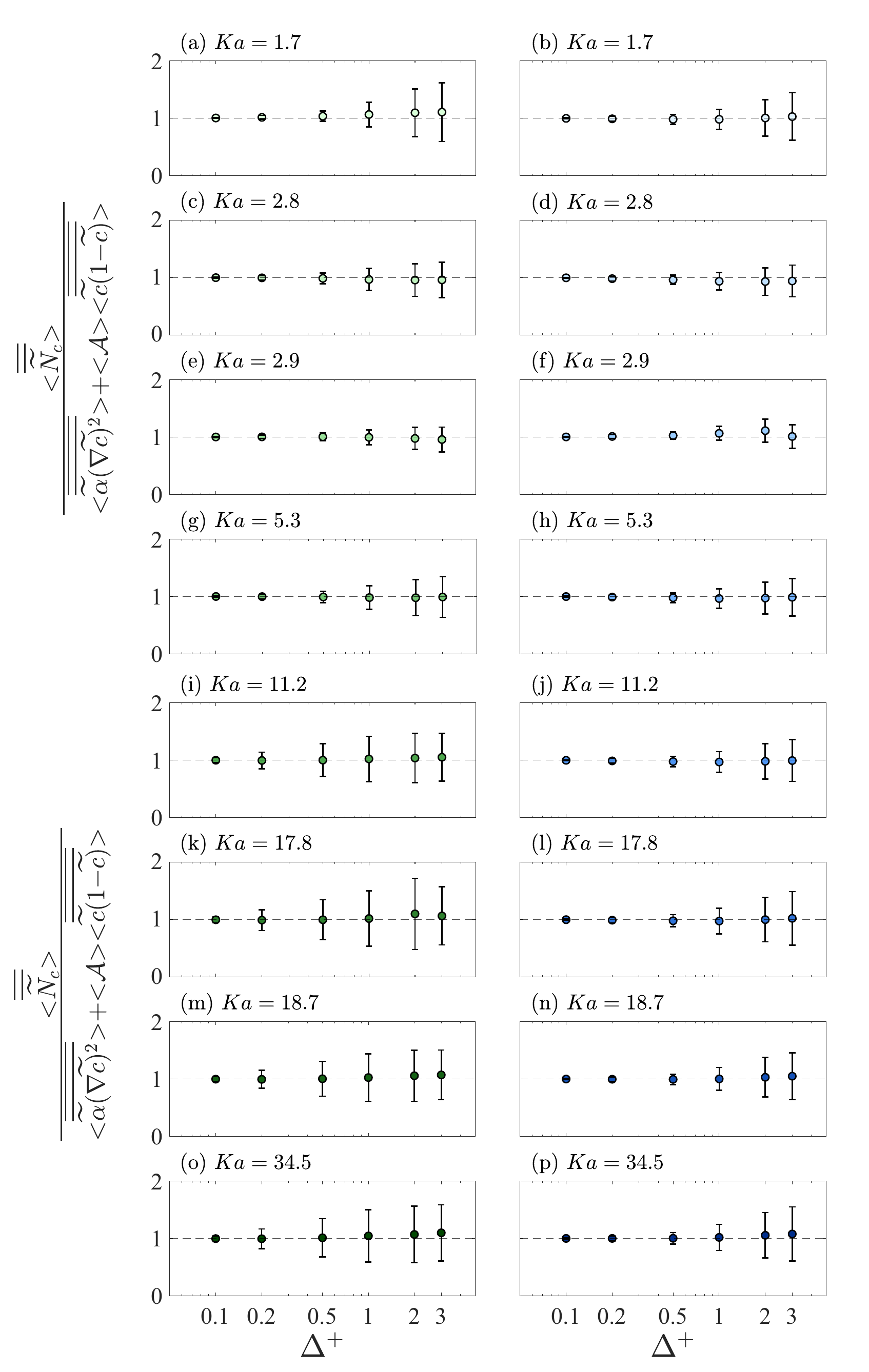}}
\caption{Variation of the ratio of $<\overline{\overline{\widetilde{N_c}}}>$ and $<\overline{\overline{\widetilde{\alpha}(\nabla\widetilde{c})^2}}>+<\mathcal{A}><\overline{\overline{\widetilde{c}(1-\widetilde{c})}}>$ for all examined test conditions, versus the normalized filter size for Gaussian (left column, in green) and box (right column, in blue) filtering methods. The symbols $<>$ and over double-bar represent the spatial and temporal averaging, respectively. The error bars are estimated based on the sensitivity of the ratio to the spatial variations of $\mathcal{A}$.}
\label{Fig:STDratio}
\end{figure}

The effect of the spatial variation in $\mathcal{A}$ on the accuracy of predicting the scalar dissipation rate was further investigated. The values of $<\overline{\overline{\widetilde{N_c}}}>/[<\overline{\overline{\widetilde{\alpha}(\nabla\widetilde{c})^2}}>+<\mathcal{A}><\overline{\overline{\widetilde{c}(1-\widetilde{c})}}>]$ were calculated and the standard deviation of the values were obtained, given the spatial variation of $\mathcal{A}$. These standard deviations along with $<\overline{\overline{\widetilde{N_c}}}>/[<\overline{\overline{\widetilde{\alpha}(\nabla\widetilde{c})^2}}>+<\mathcal{A}><\overline{\overline{\widetilde{c}(1-\widetilde{c})}}>]$ are presented in Fig.~\ref{Fig:STDratio} using error bars and solid symbols, respectively. Small variations (less than 5\% of $<\overline{\overline{\widetilde{N_c}}}>/[<\overline{\overline{\widetilde{\alpha}(\nabla\widetilde{c})^2}}>+<\mathcal{A}><\overline{\overline{\widetilde{c}(1-\widetilde{c})}}>]$) are reported for test conditions with Karlovitz numbers smaller than 5.3 and normalized filter sizes smaller than 0.5 for both Gaussian and box filtering methods. For these normalized filter sizes, the deviation remains nearly unchanged for the box filtering method and for $Ka > 5.3$. This is consistent with the results presented in Fig.~\ref{Fig:Aspace}. Increasing the filter size and $Ka$ increases the deviation. The largest deviation corresponds to $Ka = 17.8$ and $\Delta^+ = 2.0$ ($\Delta^+ = 3.0$), with the error bar being about 56\% (45\%) of $<\overline{\overline{\widetilde{N_c}}}>/[<\overline{\overline{\widetilde{\alpha}(\nabla\widetilde{c})^2}}>+<\mathcal{A}><\overline{\overline{\widetilde{c}(1-\widetilde{c})}}>]$ for the Gaussian filtering (box filtering) method. Such large error bars at large $\Delta^+$ is linked to the relatively large variation of $<\mathcal{A}>$ as presented in Fig.~\ref{Fig:Aspace}. In essence, the analyses in this appendix quantifies the effect of $\mathcal{A}$ spatial variation on predicting $<\overline{\overline{\widetilde{N_c}}}>$. Ultimately, the present study used $<\mathcal{A}>$ for predicting $<\overline{\overline{\widetilde{N_c}}}>$.

\clearpage

\bibliography{sn-bibliography}

\end{document}